%% file: main.tex
\documentclass[aps,prl,letterpaper,twocolumn,superscriptaddress,showpacs,amsmath,amssymb,nofootinbib,floatfix]{revtex4-1}
\usepackage{graphicx}  % needed for figures
\graphicspath{ {figures/} } 
\usepackage{bm}     % for math
\usepackage{amssymb}   % for math
\usepackage{amsmath}   % for math
\usepackage{hyperref}% add hypertext capabilities
\usepackage{todonotes}
\usepackage[export]{adjustbox}
\usepackage{textcomp}
\usepackage[version=4]{mhchem}
\usepackage{siunitx}
\usepackage{multirow}
\usepackage{subfigure}
\usepackage{xspace}
\usepackage{soul}
\usepackage{mathtools}
\DeclarePairedDelimiter\set\{\}

\input{commands.tex}

\begin{document}
\title{First Measurement of Coherent Elastic Neutrino-Nucleus Scattering on Argon}
\input{authors.tex}  
\date{\today}

\begin{abstract}
We report the first measurement of coherent elastic neutrino-nucleus scattering (\cevns) on argon using a liquid argon detector at the Oak Ridge National Laboratory Spallation Neutron Source. Two independent analyses prefer \cevns over the background-only null hypothesis with greater than $3\sigma$ significance. The measured cross section, averaged over the incident neutrino flux, is (2.2 $\pm$ 0.7) $\times$10$^{-39}$ cm$^2$ --- consistent with the standard model prediction. The neutron-number dependence of this result, together with that from our previous measurement on CsI, confirms the existence of the \cevns process and provides improved constraints on non-standard neutrino interactions.
\end{abstract}
\maketitle
 
%(Word count: \quickwordcount{main})
\paragraph{\label{sec:intro}Introduction ---} 
Coherent elastic neutrino-nucleus scattering (\cevns)~\cite{freedman74,Kopeliovich:1974mv} occurs when a neutrino interacts coherently with the total weak nuclear charge, necessarily at low momentum transfer, leaving the ground state nucleus to recoil elastically. It is the dominant interaction for neutrinos of energy $E_\nu \lesssim 100$~MeV and provides a sensitive test of standard model (SM) and beyond-SM processes~\cite{Barranco:2005yy,Barranco:2007tz,Dutta:2015vwa,Krauss:1991ba}. 

In this Letter, we report the first measurement of \cevns in a light nucleus (argon) complementing our earlier result on cesium and iodine~\cite{Akimov:2017ade}, thus establishing the $N^{2}$ behavior predicted by the standard model.  This result also improves constraints on non-standard interactions between neutrinos and quarks. 

\cevns is sensitive to these non-standard interactions (NSI), which are crucial to understand for the success of the long-baseline neutrino oscillation program~\cite{Coloma:2016gei,Coloma:2017egw,Coloma:2017ncl,Marfatia16}.  The process also probes the weak nuclear charge~\cite{Amanik:2009zz,Cadeddu:2018izq,Patton:2012jr,Cadeddu:2017etk,AristizabalSierra:2019zmy,Hoferichter:2018acd} and the weak mixing angle at novel momentum transfer~\cite{Krauss:1991ba,canas2018}. Additionally, \cevns-sensitive detectors could play future roles as non-intrusive nuclear reactor monitors~\cite{Barbeau2002,Hagmann2004,KIM2016285}.

\cevns has numerous connections to possible hidden-sector particles. It is sensitive to $Z'$ models which could explain the theoretical tension with measurements of the muon anomalous magnetic moment~\cite{davoudiasl2014}. \cevns from solar and atmospheric neutrinos constitute the so-called ``neutrino floor'' background in future dark matter searches~\cite{Gonzalez-Garcia:2018}, and \cevns cross section measurements quantify this background. \cevns experiments at accelerators are also sensitive to sub-GeV accelerator-produced dark matter particle models~\cite{deNiverville:2015mwa,Ge:2017mcq,Dutta:2019nbn,
Dutta:2019,Akimov:2019xdj}. The potential relevance of \cevns to core-collapse supernovae was quickly recognized~\cite{Freedman:1977}, and though its role in supernova dynamics is uncertain~\cite{Janka:2012,balasi2015}, \cevns is expected to be the source of neutrino opacity in these events~\cite{bruenn1997}. Supernova neutrinos convey information about supernova dynamics, and could be detected via \cevns~\cite{Horowitz:2003}. 

\cevns measurements require detectors with low nuclear-recoil-energy threshold in a low-background environment with an intense neutrino flux. The COHERENT collaboration has deployed a suite of detectors in a dedicated neutrino laboratory (``Neutrino Alley'') at the Spallation Neutron Source (SNS) at Oak Ridge National Laboratory (ORNL)~\cite{Akimov:2017ade,Akimov:2018ghi}. We reported the first observation of \cevns on heavy nuclei using a 14.6-kg, low-background, low-threshold \csina detector located 19.3~m from the SNS target~\cite{Akimov:2017ade}. 

\begin{figure}[htbp]
\includegraphics[width=0.85\columnwidth,trim={55 60 30 50},clip]{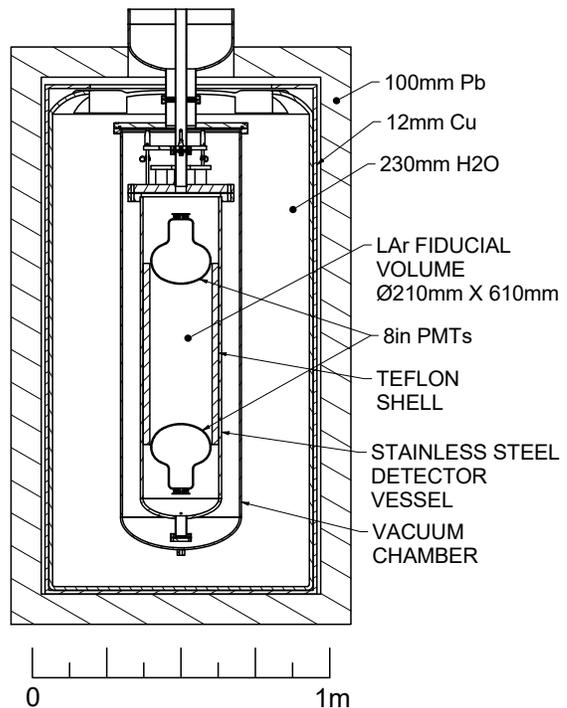}
\caption{\label{fig:CENNS10Run1} \cenns liquid argon detector and associated shielding as configured for the results reported here.}
\end{figure}

As part of the COHERENT program, we deployed the 24-kg active-mass liquid-argon (LAr) \cenns scintillator detector (Fig.~\ref{fig:CENNS10Run1}) in Neutrino Alley to detect \cevns in a light nucleus. The initial \cenns deployment set a limit on the \cevns cross section for argon and quantified backgrounds~\cite{Akimov:2019}. A subsequent upgrade provided a lower energy threshold with an eight-fold improvement in light collection efficiency.

\paragraph{\label{sec:exp}Experiment ---} 
The 1-GeV, 1.4-MW proton beam of the SNS accelerator strikes a liquid-\ce{Hg} target in $\SI{360}{\ns}$ FWHM pulses at \SI{60}{\Hz} to produce neutrons that are moderated and delivered to experiments. Additionally, $(9.0\pm0.9)\times10^{-2}$~\piplus are produced for each proton-on-target (POT) leading to a large flux of pion-decay-at-rest neutrinos. The \piplus produce a prompt  \SI{29.8}{\MeV} \numu along with a \muplus, which subsequently decays  yielding a three-body spectrum of \numubar[] and \nue[] with an endpoint energy of \SI{52.8}{\MeV}. This time structure is convolved with the proton beam pulse yielding a prompt \numu neutrino flux followed by a delayed flux of \numubar[] and \nue[]~\cite{Akimov:2017ade,Akimov:2018ghi}. 

The \cenns detector, designed and built at Fermilab~\cite{Brice:2013fwa}, sits 27.5~m from the SNS target in Neutrino Alley. The active volume of \cenns is defined by a cylindrical polytetrafluoroethylene (PTFE) shell and two 8" Hamamatsu R5912-02MOD photomultiplier tubes (PMTs) resulting in active mass of 24~kg of atmospheric argon (99.6\% \iso{40}{Ar}). The PTFE and PMT glass are coated with a 0.2~mg/cm$^2$ layer of 1,1,4,4-tetraphenyl-1,3-butadiene (TPB) to wavelength-shift the 128-nm argon scintillation light to a distribution peaked at 420~nm where the PMTs have quantum efficiency of $18$\%. This configuration provides a $\sim20$~keVnr (nuclear-recoil) energy threshold.  

Argon scintillation light from particle interactions is produced from both ``fast'' singlet ($\tau_{s}\approx 6$ ns) and ``slow'' triplet ($\tau_{t}\approx 1600$ ns) excited molecular states~\cite{Hitachi:1983zz}. Electron recoils (ER) and argon nuclear recoils (NR) populate these states in different proportions, allowing for pulse-shape discrimination (PSD) to suppress ER backgrounds from electron-gamma background sources compared to the \cevns NR recoil signal.  Neutron sources, from the accelerator or surrounding materials, will also create a NR signal, so shielding is required to reduce this background.

During SNS operation, each PMT waveform is digitized at 250~MHz in a 33-\us window around each POT pulse (``on-beam'' data) together with a subsequent 33-\us window between POT pulses (``off-beam'' data) to allow a measure of beam-unrelated backgrounds. Calibration data were acquired using \iso{57}{Co} and \iso{241}{Am} sources placed within the water shield, a sample of $^{83m}$Kr gas injected via the argon re-circulation system~\cite{Krpaper}, as well as an external americium-beryllium (AmBe) neutron source. A pulsed visible-spectrum LED, along with triplet light from low light-yield calibration pulses, was used to determine the response of the PMTs to single-photoelectron (SPE) signals. These calibration runs were performed on a weekly basis to correct for drifts in detector response due to PMT gain or light output changes.

\paragraph{Analysis ---} 
In order to avoid experimenter bias, the analysis methods and event selection criteria were established, prior to examining the on-beam data set, by two independent analysis groups — labeled as ``A'' and ``B'' below.

The PMT waveforms were integrated over a 6~$\mu$s window after the initial PMT pulse and summed to form the integrated event amplitude, $I$.  Also, the integrated amplitude in the first $90$~ns, $I_{90}$, was calculated and the PSD parameter $\fninety=I_{90}/I$ defined. Off-beam and on-beam windows were treated identically, providing an unbiased measurement of the beam-unrelated backgrounds. The $\gamma$-ray sources were used to calibrate scintillation yield to electron-equivalent energy (keVee) with $2$\% uncertainty. The energy resolution was 9\% at the $41.5$~keVee $^{83m}$Kr line. A comparison of the calibration source signals to SPE signals from a pulsed LED and from delayed low-light-yield events resulted in an estimated $\sim4.5$~photoelectrons (PE) per keVee. 

The detector response to \cevns\ NR events compared to calibration ER events is quantified via the so-called ``quenching factor'' ($\mathrm{QF}$).  We performed a linear fit to the world data~\cite{Agnes:2018mvl,Cao:2014gns,Creus:2015fqa, Gastler:2010sc} for QF on argon in the energy range $0-125$~keVnr following the Particle Data Group prescription for combining measurements~\cite{PDG2018}, incorporating the correlated uncertainties reported in Ref.~\cite{Gastler:2010sc}. With this fit and the ER calibration from above, the response to \cevns\ NR events can be simulated. At 20~keVnr, the fit yields  $\mathrm{QF}=0.26\pm0.01$.  The AmBe neutron source data were used to determine the PSD response for NR events via the use of the quantity $\fninety$ with energy dependence consistent with other measurements in LAr~\cite{Hitachi:1983zz,Regenfus_2012}.

A \textsc{Geant4}-based~\cite{Agostinelli:2002hh} program modeled the detector response for both \cevns and neutron events to determine the \cevns detection efficiency and construct predicted event distributions. The program simulates the production and quenching of LAr scintillation light, TPB absorption and re-emission, and propagation of optical photons to the PMTs. The material optical parameters and LAr scintillation properties were adjusted to reproduce the calibration data and then used to estimate the \cevns response and detection efficiency.

The beam-unrelated ``steady-state'' (SS) background was measured \textit{in situ} using the off-beam triggers occuring one-for-one with on-beam triggers.  The time window within the off-beam trigger can be made larger that the on-beam time window, allowing for an ``oversampling'' of the background, thus reducing the systematic uncertainty on the measured rate to $<1\%$. In addition, the energy and $\fninety$ distributions are also precisely measured, eliminating the need for knowledge of the exact source of this background and for any additional systematic errors.  Qualitatively, the measured spectrum is consistent with a dominant background from the 565~keVee-endpoint $\beta$-decay of \iso{39}{Ar} in the detector volume. The remainder is mostly from $\gamma$-rays from surrounding materials or a nearby SNS target radioactive gas exhaust pipe, which are suppressed by the Pb shielding. Relative to the on-beam signal, these backgrounds are $\sim10^4$-fold suppressed due to the pulsed SNS beam structure and $\sim10^2$-fold further suppressed by PSD in the event selection.

The beam-related background events are caused by neutrons originating in the SNS target that elastically scatter in the argon, producing a NR event. Though this beam-related neutron (BRN) rate is highly suppressed in Neutrino Alley, the events occur in time with the beam, and the rate competes with the \cevns rate in the detector. The BRN flux at the \cenns location was measured with the SciBath neutron detector~\cite{Tayloe:2006ct,Cooper:2011kx} in 2015, was further studied with the \cenns engineering run~\cite{Akimov:2019}, and was measured as part of this analysis in a three-week (0.54~GW$\cdot$hr) ``no-water'' run in which the water shielding around the detector was drained. Neutrino-induced neutrons from neutrino interactions in the lead shielding~\cite{kolbe2001} can also produce prompt NR events; however, the water shielding between the lead and detector reduces their contribution to $<$ 1 event in this data set.

\begin{figure}[htbp]
\includegraphics[width=0.99\columnwidth,trim={10 0 50 0},clip]{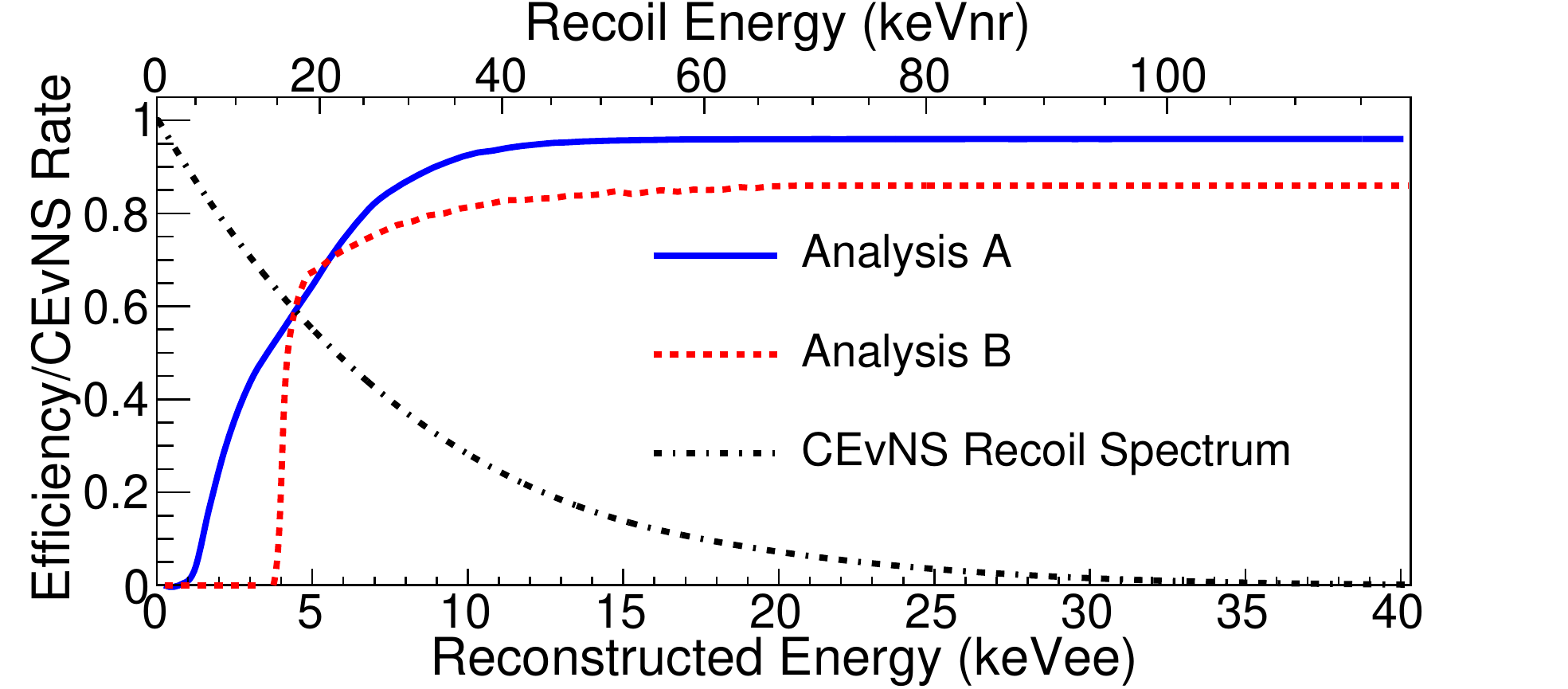}
\caption{\label{fig:eff} Energy-dependent \cevns event efficiency for analyses A (blue) and B (red dash) along with the predicted \cevns recoil spectrum in arbitrary units (black dash-dot). Analysis B requires $E>4.1$~keVee, as well as a ``top-fraction'' cut (see text).}
\end{figure}

The data used for this \cevns\ analysis correspond to total integrated beam power of 6.12~GW$\cdot$hr  (13.7 $\times$ 10$^{22}$~POT) collected between July 2017 -- December 2018. 
Events are selected from both on-beam and off-beam data sets with identical cuts.  Candidate events are initiated by requiring pulses with $\ge2$~PE in both PMTs occurring within $20$~ns of each other. This cut largely determines the energy threshold and rejects 15\% of the predicted \cevns events at lowest recoil energies. Pulses within an event must not exhibit preceding or delayed ``pileup'' pulses, rejecting a further 4\% of events. In addition, Analysis B required that each PMT recorded at least 20\% of the total light in an event, reducing some background events that occur near either PMT  while reducing the \cevns event selection efficiency by 10\%.  

Further, candidate events were required to lie in the $\fninety$ NR band to reject ER- and Cherenkov-like events. A time range was chosen using $t_{\textrm{trig}}$, where $t_{\textrm{trig}}=0$ is the expected start time of the neutrino beam at the detector, to include both prompt and delayed neutrinos. An energy range was chosen to include the region of interest for a \cevns signal ($E<120$~keVnr~$\approx30$~keVee). The specific values for the fit ranges, summarized in Table~\ref{tab:llresults}, differed between analyses A and B because of different strategies for signal and background optimization. For example, Analysis A used a wider energy range to include more high-energy BRN events to anchor that background so an extra delayed-BRN component would be better constrained. The resulting energy-dependent efficiency for detecting \cevns is shown in Fig.~\ref{fig:eff}.

%\rnote{could get rid of some \hlines in this?}
\begin{table}[htbp]
   \caption{Summary of parameters, errors, and results for the maximum likelihood fit and cross section extraction. Analysis A divides the BRN component into ``prompt'' and ``delayed'' parts.  ``BRN'' and ``SS'' are the beam-related-neutron and steady-state backgrounds.}
\centering
    \begin{tabular}{l|rcr|rcr}
        \hline\hline       
% FIT RANGES         
       \textbf{fit ranges} & \multicolumn{3}{c|}{Analysis A} & \multicolumn{3}{c}{Analysis B} \\
        \hline
        $\fninety$ & 0.5 & $-$ & 0.9 & 0.5 & $-$ & 0.8 \\
       $E$ (keVee)  & 0.0 & $-$ & 120.0 & 4.1 & $-$ & 30.6 \\  
        $t_\textrm{trig}$ ($\mu$s)   & $-0.1$ & $-$ & $4.9$ & $-1.0$ & $-$ & 8.0 \\
        \hline
        total events selected & \multicolumn{3}{c|}{3752} & \multicolumn{3}{c}{1466} \\
        \hline
% FIT RANGES (alternative to have less \hlines)        
%        & \multicolumn{3}{c|}{Analysis A} & \multicolumn{3}{c}{Analysis B} \\
%        \hline
%        \multicolumn{7}{l}{\textbf{fit ranges}} \\
%        \hspace{3mm}$\fninety$ & 0.5 & $-$ & 0.9 & 0.5 & $-$ & 0.8 \\
%        \hspace{3mm}$E$ (keVee)  & 0.0 & $-$ & 120.0 & 4.1 & $-$ & 30.6 \\  
%        \hspace{3mm}$t_\textrm{trig}$ ($\mu$s)   & $-0.1$ & $-$ & $4.9$ & $-1.0$ & $-$ & 8.0 \\
%        total events selected & \multicolumn{3}{c|}{3752} & \multicolumn{3}{c}{1466} \\
%        \hline
% EVENTS PREDICTED
        \multicolumn{7}{l}{\textbf{input values}} \\
%       \multicolumn{7}{l}{\textbf{predicted}} \\
        \hline
        $\Ncevns$         & 128 & $\pm$ & 17 & 101 & $\pm$ & 12 \\
        $\Nbrn$, prompt   & 497 & $\pm$ & 160 & \multirow{2}{*}{226} & \multirow{2}{*}{$\pm$} & \multirow{2}{*}{33} \\
        $\Nbrn$, delayed   & 33 & $\pm$ & 33 & & & \\
        $\Nss$            & 3152 & $\pm$ & 25 & 1155 & $\pm$ & 45 \\
        \hline
        total events predicted & \multicolumn{3}{c|}{3779} & \multicolumn{3}{c}{1482}  \\
        \hline
% EVENTS FIT
        \multicolumn{7}{l}{\textbf{fit values} } \\
        \hline
        $\Ncevns$ & 159 & $\pm$ & 43 & 121 & $\pm$ & 36 \\
        $\Nbrn$, prompt    & 553 & $\pm$ & 34 & \multirow{2}{*}{222} & \multirow{2}{*}{$\pm$} & \multirow{2}{*}{23} \\
        $\Nbrn$, delayed   & 10 & $\pm$ & 11 & &  & \\
        $\Nss$    & 3131 & $\pm$ & 23 & 1112 & $\pm$ & 41 \\
        \hline
        total events fit & \multicolumn{3}{c|}{3853} & \multicolumn{3}{c}{1455}  \\
        \hline  
% SYSTEMATICS        
        \multicolumn{7}{l}{\textbf{fit systematic errors}} \\
        \hline    
        \cevns $\fninety$ $E$ dependence       & \multicolumn{3}{c|}{4.5\%} & \multicolumn{3}{c}{3.1\%} \\ 
        \cevns $t_\textrm{trig}$ mean        & \multicolumn{3}{c|}{2.7\%} & \multicolumn{3}{c}{6.3\%} \\ 
        BRN $E$ dist.                        & \multicolumn{3}{c|}{5.8\%} & \multicolumn{3}{c}{5.2\%} \\ 
        BRN $t_\textrm{trig}$ mean           & \multicolumn{3}{c|}{1.3\%} & \multicolumn{3}{c}{5.3\%} \\ 
        BRN $t_\textrm{trig}$ width          & \multicolumn{3}{c|}{3.1\%} & \multicolumn{3}{c}{7.7\%} \\ 
        \hline        
        total \cevns sys. error              & \multicolumn{3}{c|}{8.5\%} & \multicolumn{3}{c}{13\%} \\ 
        \hline
% FIT RESULTS    
        \multicolumn{6}{l}{\textbf{fit results}} \\
        \hline
        null significance (stat. only)  & \multicolumn{3}{c|}{3.9$\sigma$} & \multicolumn{3}{c}{3.4$\sigma$} \\
        null significance (stat.+sys.)& \multicolumn{3}{c|}{3.5$\sigma$} & \multicolumn{3}{c}{3.1$\sigma$} \\
        \hline
  % CROSS SECTION  
        \multicolumn{7}{l}{\textbf{cross section}} \\
        \hline
        SM-predicted $\sigma$ ($\times10^{-39}$~cm$^2$)  & \multicolumn{6}{c}{1.8} \\ 
        \hline
        \multicolumn{7}{l}{systematic errors:} \\ \hline
        detector efficiency  & \multicolumn{3}{c|}{3.6\%} & \multicolumn{3}{c}{1.6\%} \\    
        energy calibration   & \multicolumn{3}{c|}{0.8\%} & \multicolumn{3}{c}{4.6\%} \\    
        \fninety calibration & \multicolumn{3}{c|}{7.8\%} & \multicolumn{3}{c}{3.3\%} \\  
        quenching factor     & \multicolumn{3}{c|}{1.0\%} & \multicolumn{3}{c}{1.0\%} \\ 
        nuclear form factor  & \multicolumn{3}{c|}{2.0\%} & \multicolumn{3}{c}{2.0\%} \\
        neutrino flux        & \multicolumn{3}{c|}{10\%} & \multicolumn{3}{c}{10\%} \\ \hline      
        total cross section sys. error & \multicolumn{3}{c|}{13\%} & \multicolumn{3}{c}{12\%} \\ \hline
        measured $\sigma$ ($\times10^{-39}$~cm$^2$) & \multicolumn{3}{c|}{2.3 $\pm$ 0.7} & \multicolumn{3}{c}{2.2 $\pm$ 0.8} \\
        \hline\hline  
    \end{tabular}
    \label{tab:llresults}
\end{table}

\begin{figure*}[htbp]
\begin{minipage}{0.325\textwidth}
\includegraphics[width=\textwidth,trim={0 1 40 0},clip]{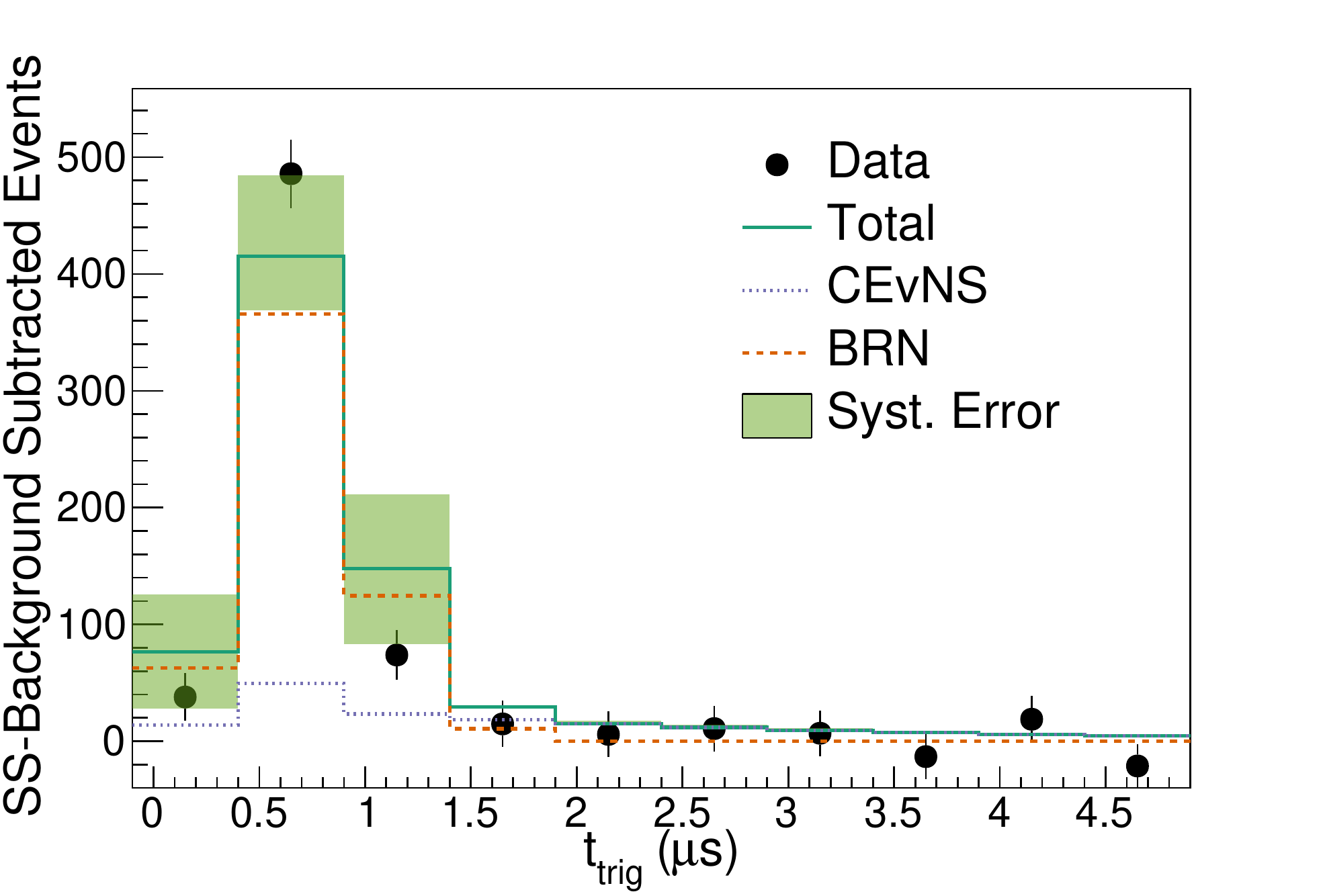}
\end{minipage}
\hfill
\begin{minipage}{0.325\textwidth}
\includegraphics[width=\textwidth,trim={0 1 40 0},clip]{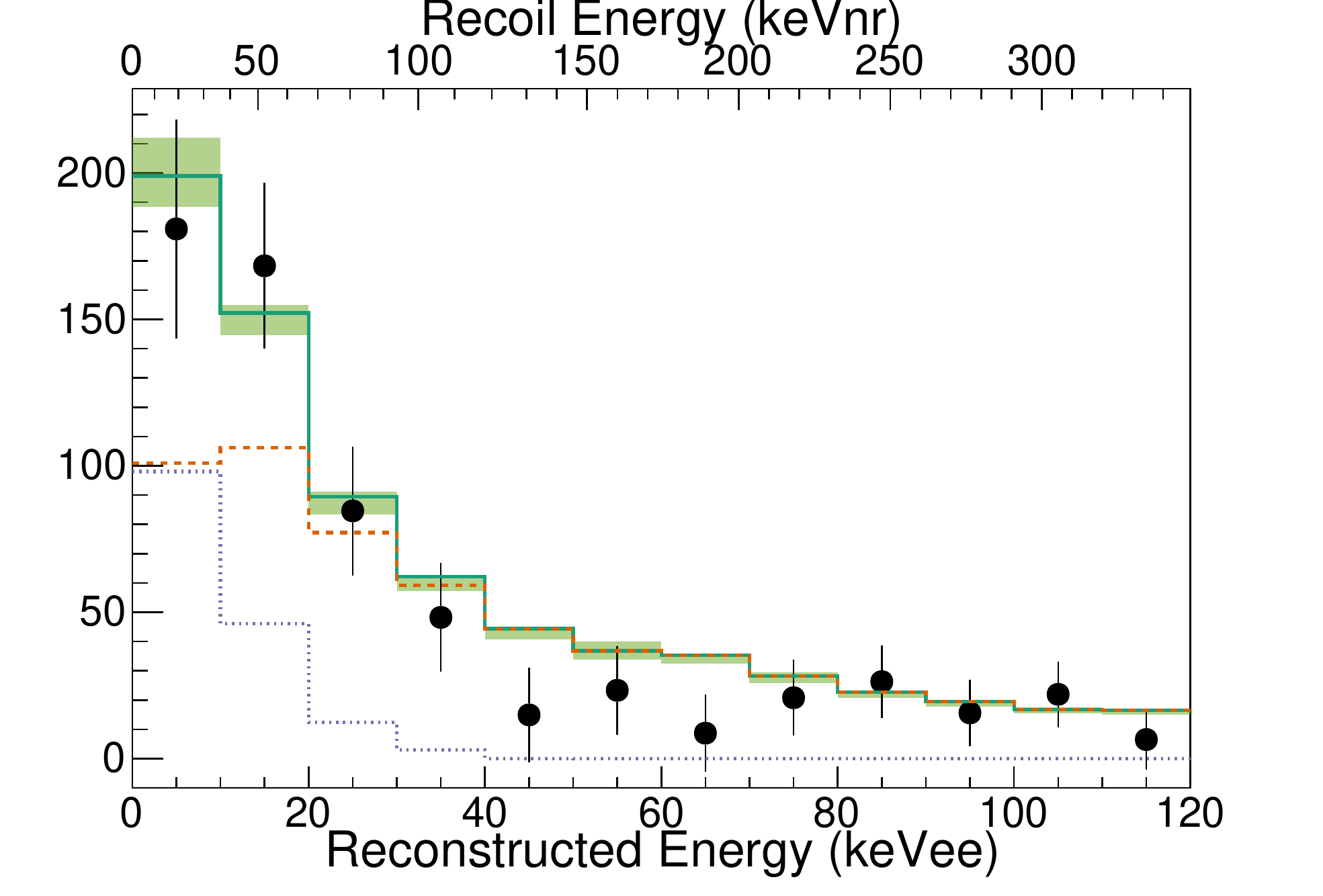}
\end{minipage}
\hfill
\begin{minipage}{0.325\textwidth}
\includegraphics[width=\textwidth,trim={0 1 40 0},clip]{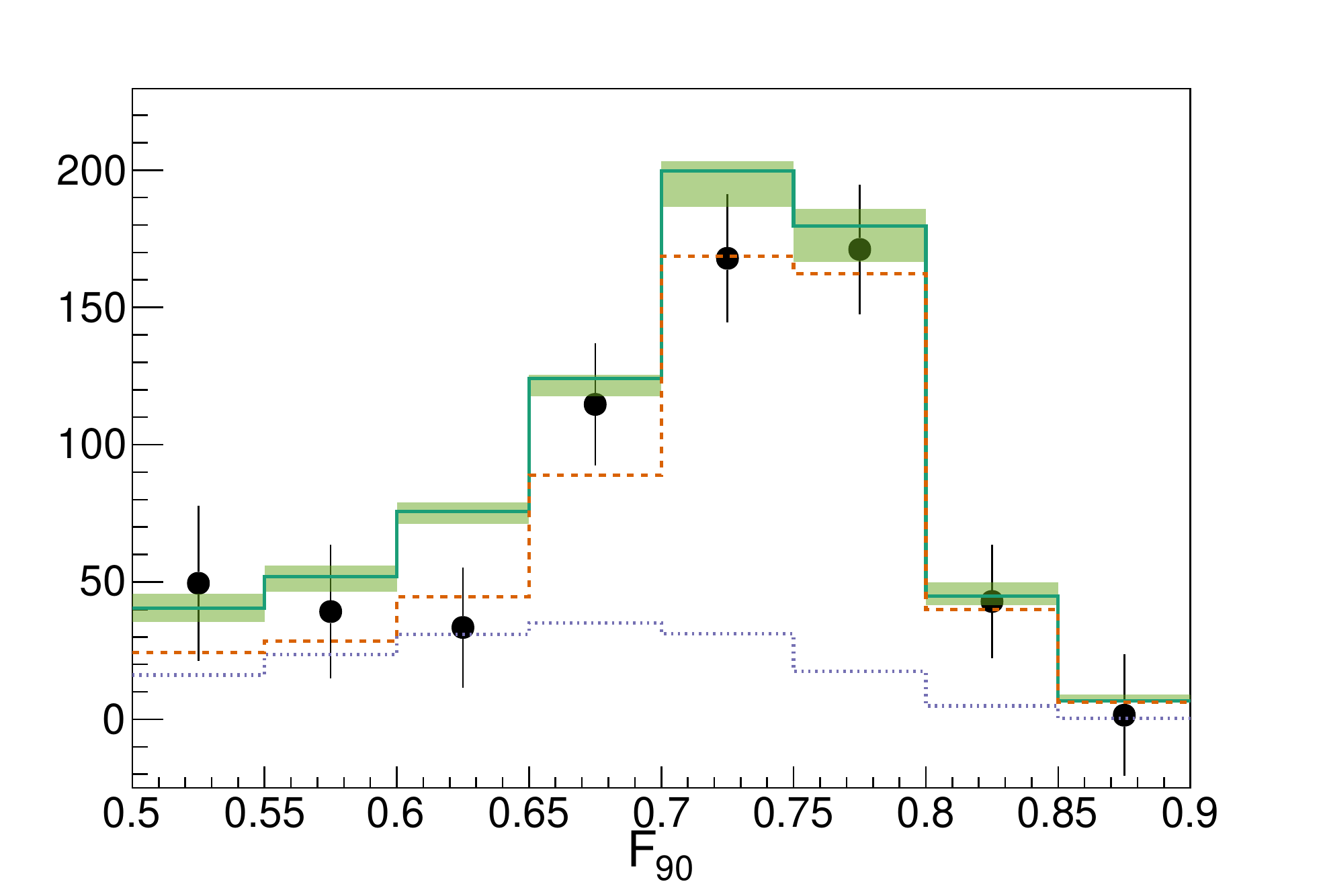}
\end{minipage}
\caption{\label{fig:llprojections} Projection of the best-fit maximum likelihood probability density function (PDF) from Analysis A on $t_\mathrm{trig}$ (left), reconstructed energy (center), and $\fninety$ (right) along with selected data and statistical errors. The fit SS background has been subtracted to better show the \cevns component. The green band shows the envelope of fit results resulting from the $\pm 1\sigma$ systematic errors on the PDF.}
\end{figure*}

For the extraction of \cevns events amid BRN and SS backgrounds, we performed an extended maximum-likelihood fit to the on-beam data binned in $\fninety$, $t_{\mathrm{trig}}$, and $E$. These data were modeled by distributions $\mathcal{P}_{k}\left(E,t_{\mathrm{trig}},\fninety \right)$ with associated number of events $N_{k}$ for $k\in\set{\mathrm{\cevns},\mathrm{BRN},\mathrm{SS}}$. The best-fit number of \cevns events, $\Ncevns$, was unconstrained in the fit.
$\Pcevns$ was determined from a simulation of \cevns events to provide the PSD and energy distributions, then combining with the neutrino arrival-time dependence. 

For the backgrounds, the total number of SS events, $\Nss$, was Gaussian-constrained by the statistical error from the off-beam measurement of $0.8\%$ ($3.8\%$) for Analysis A (B). The $\Pss$ distribution was formed by binning the off-beam events in $E$ and $\fninety$ and assuming a constant time dependence. Analysis A Gaussian-constrained $\Nbrn$ based on associated BRN measurements; Analysis B allowed $\Nbrn$ to float freely.  Analysis A also included a separate delayed ($1.4<t_\mathrm{trig}<1.9$~$\mu$s) BRN component in the fit to permit the possibility of late BRN events. For $\Pbrn$, the $\fninety$-$E$-dependence was extracted from the simulation with a time dependence extracted from a fit to the no-water data. 

Pseudo-data sets were generated using RooFit~\cite{Verkerke:2003ir} to demonstrate a robust and unbiased fitting procedure, and to estimate uncertainties before fitting the on-beam data.  Only systematic uncertainties that affect the shape of the $\mathcal{P}_{k}$ affect the fit value of $\Ncevns$. The individual contributions are treated as independent and added in quadrature for the total systematic error on the fit number of \cevns\ events.  
\paragraph{Results ---} 
The input parameters, errors, and results for the maximum likelihood fit of $\Ncevns$ for both analyses are summarized in Table~\ref{tab:llresults}. The significance of this result compared to the null hypothesis, incorporating systematic errorsas explained above, is $3.5\sigma$ ($3.1\sigma$) for Analysis A (B). Both analyses yield $\Ncevns$ within 1$\sigma$ of the SM prediction. Note that the large SS background is not as detrimental to signal significance as expected with a simple signal to background argument because it is well-measured and of different character than signal in the $\mathcal{P}_{k}\left(E,t_{\mathrm{trig}},\fninety \right)$ distributions.

The data and best fit for analysis A are shown in Fig.~\ref{fig:llprojections}, projected along $E$, $\fninety$, and $t_{\mathrm{trig}}$. Extraction of the relatively low-energy \cevns signal is robust in the presence of the large prompt BRN background because of the latter's much harder spectrum. 

\begin{figure}[htbp]
\resizebox{0.99\columnwidth}{6cm}{\includegraphics[trim={0 0 50 30},clip]{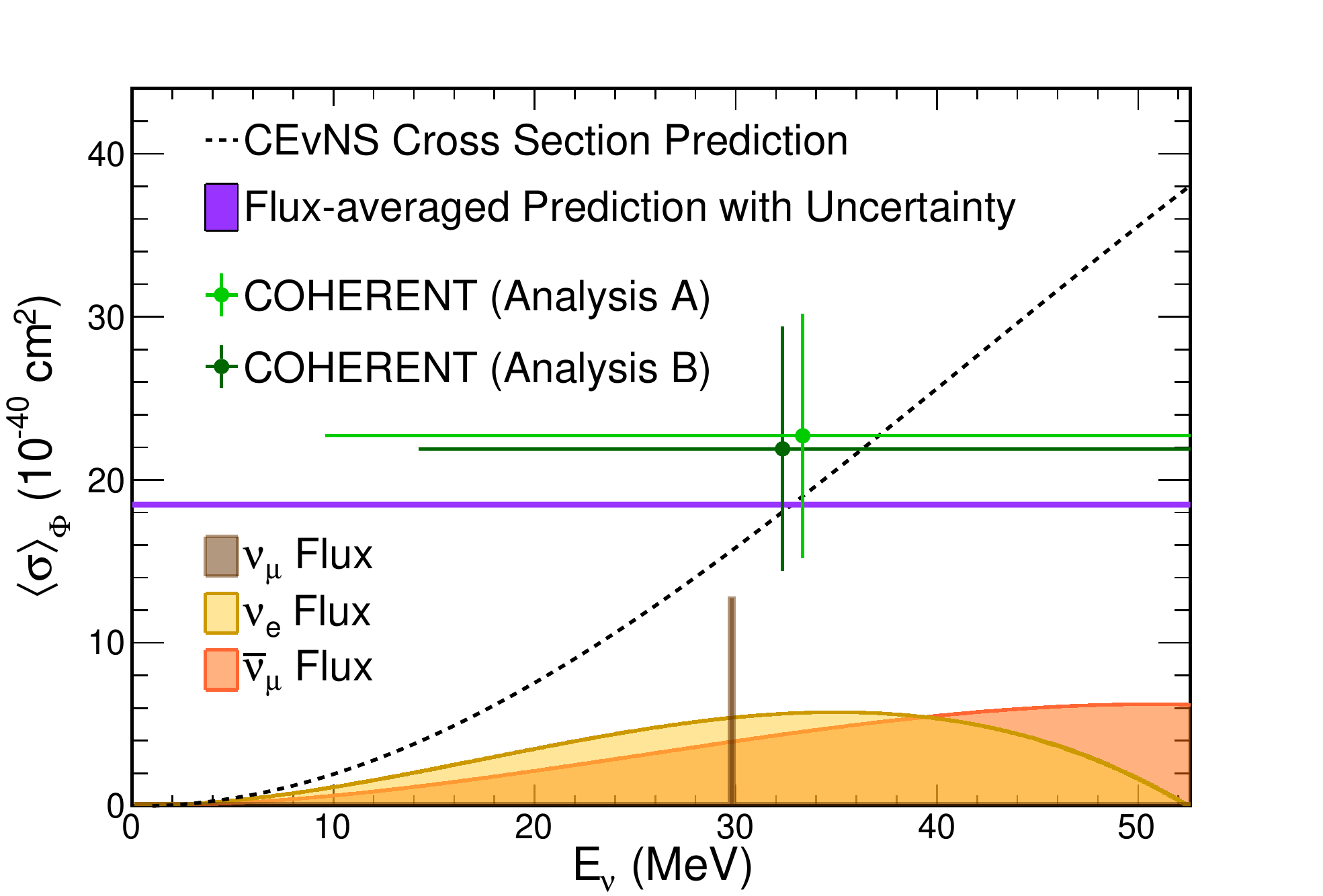}}
\caption{\label{fig:xsectionvE} Measured \cevns flux-averaged cross section for the two analyses, along with the SM prediction.  The horizontal bars indicate the energy range of the flux contributing.  The minimum value is set by the NR threshold energy and is different for each analysis. The 2$\%$ error on the theoretical cross section due to uncertainty in the nuclear form factor is also illustrated by the width of the band. The SNS neutrino flux is shown with arbitrary normalization.}
\end{figure}

We compute the \cevns flux-averaged cross section on argon (99.6\% \iso{40}{Ar}) from the ratio of the best-fit $\Ncevns$ to that predicted by the simulation using the SM prediction of $1.8\times10^{-39}$~cm$^2$. This incorporates the total uncertainty on the fit $\Ncevns$ along with additional systematic uncertainties, dominated by the 10\% incident neutrino flux uncertainty, that do not affect the signal significance. The values are summarized along with extracted cross section values in Table~\ref{tab:llresults}. The measured flux-averaged cross sections are consistent between the two analyses and with the SM prediction as shown in Fig.~\ref{fig:xsectionvE}. We average the results of the two analyses to obtain $(2.2\pm0.7)\times10^{-39}$~cm$^2$ with uncertainty dominated by the $\sim 30$\% statistical uncertainty on \Ncevns.

\begin{figure}[htbp]
\includegraphics[width=0.99\columnwidth,trim={10 10 10 10},clip]{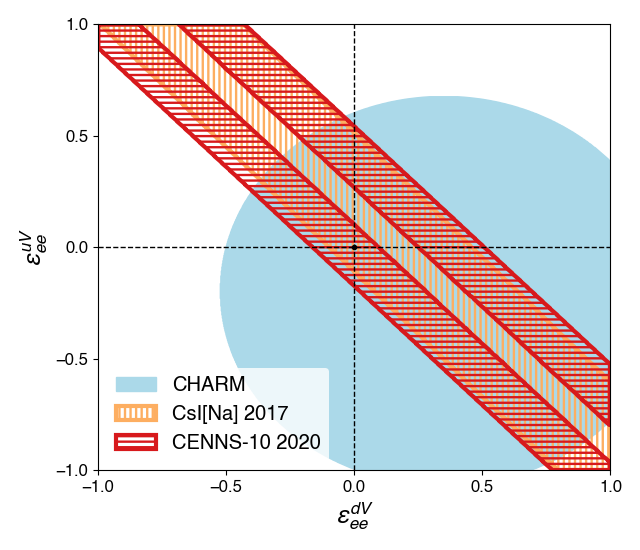}
\caption{\label{fig:nsi} 90\% CL regions for non-standard neutrino interactions (NSI) for a vector-coupled quark-electron interaction extracted from this argon measurement plotted together with the previous COHERENT CsI[Na] measurement~\cite{Akimov:2017ade} and the CHARM experiment~\cite{Dorenbosch:1986tb}.  The 3 regions shown are independent and the dashed black lines show the SM prediction. 
}
\end{figure}

This result is used to constrain neutrino-quark NSI mediated by a new heavy vector particle using the framework developed in Refs.~\cite{Barranco:2005yy,Coloma:2017ncl}. Here we consider the particular case of non-zero vector-like quark-$\nu_e$ NSI couplings, $\epsilon^{uV}_{ee}$ and $\epsilon^{dV}_{ee}$, as these two are the least experimentally constrained.  The other couplings in this framework~\cite{Coloma:2017egw} are assumed to be zero.  A comparison of the measured \cevns cross section reported here to the predicted cross section including these couplings is used to determine the 90\% CL (1.65 $\sigma$) regions of NSI parameters as shown in Fig.~\ref{fig:nsi}.  The same procedure was separately applied using our previous CsI[Na] result~\cite{Akimov:2017ade} and also plotted in Fig.~\ref{fig:nsi}. The Ar measurement, with a slight excess over the SM prediction, favors a slightly different region than CsI[Na] and results in a bifurcated region because the central area corresponds to values of $\epsilon^{uV}_{ee}$ and $\epsilon^{dV}_{ee}$ that yield a cross section somewhat less than the SM value. The data and predicted background are available~\cite{Akimov:2020czh} for alternative fits.

\paragraph{\label{sec:concl}Summary ---} 
A 13.7 $\times$ 10$^{22}$ protons-on-target sample of data, collected with the \cenns detector in the SNS neutrino alley at 27.5~m from the neutron production target, was analyzed to measure the \cevns process on argon. Two independent analyses observed a more than 3$\sigma$ excess over background, resulting in the first detection of \cevns in argon. We measure a flux-averaged cross section of $(2.2\pm0.7)\times10^{-39}$~cm$^{2}$ averaged over and consistent between the two analyses. This is the second, and much lighter, nucleus for which \cevns has been measured, verifying the expected neutron-number dependence of the cross section and improving constraints on non-standard neutrino interactions. \cenns is collecting additional data which will provide, along with refined background measurements, more precise results in near future.

\paragraph{Acknowledgments ---} 
The COHERENT collaboration acknowledges the generous resources provided by the ORNL Spallation Neutron Source, a DOE Office of Science User Facility, and thanks Fermilab for the continuing loan of the \cenns detector. This material is based upon work supported by the U.S. Department of Energy, Office of Science, Office of Workforce Development for Teachers and Scientists, Office of Science Graduate Student Research (SCGSR) program. The SCGSR program is administered by the Oak Ridge Institute for Science and Education (ORISE) for the DOE. ORISE is managed by ORAU under contract number DE-SC0014664. We also acknowledge support from the Alfred~P. Sloan Foundation, the Consortium for Nonproliferation Enabling Capabilities, the Institute for Basic Science (IBS-R017-D1-2020-a00/IBS-R017-G1-2020-a00), the National Science Foundation, and the Russian Foundation for Basic Research (projs. 17-02-01077\_a, 20-02-00670\_a, and 18-32-00910 mol\_a). The work was supported by the Ministry of Science and Higher Education of the Russian Federation, Project “Fundamental properties of elementary particles and cosmology” No 0723-2020-0041 and the Russian Science Foundation, contract No.18-12-00135. Laboratory Directed Research and Development funds from ORNL also supported this project. This research used the Oak Ridge Leadership Computing Facility, which is a DOE Office of Science User Facility. This manuscript has been authored by UT-Battelle, LLC, under contract DE-AC05-00OR22725 with the US Department of Energy (DOE). The US government retains and the publisher, by accepting the article for publication, acknowledges that the US government retains a nonexclusive, paid-up, irrevocable, worldwide license to publish or reproduce the published form of this manuscript, or allow others to do so, for US government purposes. DOE will provide public access to these results of federally sponsored research in accordance with the DOE Public Access Plan \url{(http://energy.gov/downloads/doe-public-access-plan)}.
\bibliography{main.bib}

\input{supplemental}

\end{document}

%% file: commands.tex
\newcommand{\QF}{\ensuremath{\mathrm{QF}}}

\newcommand{\cevns}{\protect{CEvNS}\xspace}
\newcommand{\fninety}{\protect{F\ensuremath{{}_{90}}}\xspace}

\newcommand{\Pcevns}{\ensuremath{\mathcal{P}_{\mathrm{CE}\nu\mathrm{NS}}}}
\newcommand{\Ncevns}{\ensuremath{N_{\mathrm{CEvNS}}}}
\newcommand{\Pss}{\ensuremath{\mathcal{P}_{\mathrm{SS}}}}
\newcommand{\Nss}{\ensuremath{N_{\mathrm{SS}}}}
\newcommand{\Pbrn}{\ensuremath{\mathcal{P}_{\mathrm{BRN}}}}
\newcommand{\Nbrn}{\ensuremath{N_{\mathrm{BRN}}}}

\newcommand{\iso}[2]{\protect{\ensuremath{{}^{#1}\textrm{#2}}}\xspace}

\newcommand{\cenns[1]}[ ]{CENNS-10{#1}}

\newcommand{\nue}[1][ ]{\ensuremath{\nu_{e}}{#1}}
\newcommand{\numu}[1][ ]{\ensuremath{\nu_{\mu}}{#1}}

\newcommand{\numubar}[1][ ]{\ensuremath{\overline{\nu}_{\mu}}{#1}}

\newcommand{\piplus}[1][ ]{\ensuremath{\pi^{+}}{#1}}

\newcommand{\muplus}[1][ ]{\ensuremath{\mu^{+}}{#1}}

\newcommand{\csina}{CsI{[Na]}\xspace}

\newcommand{\us}{\protect{\ensuremath{\mu\text{s}}}\xspace}

%% file: authors.tex
% Started with author list from DM sensitiviy paper. http://inspirehep.net/record/1765499
\newcommand{\itep}{Institute for Theoretical and Experimental Physics named by A.I. Alikhanov of National Research Centre ``Kurchatov Institute'', Moscow, 117218, Russian Federation}
\newcommand{\mephi}{National Research Nuclear University MEPhI (Moscow Engineering Physics Institute), Moscow, 115409, Russian Federation}
\newcommand{\indiana}{Department of Physics, Indiana University, Bloomington, IN, 47405, USA}
\newcommand{\duke}{Department of Physics, Duke University, Durham, NC 27708, USA}
\newcommand{\tunl}{Triangle Universities Nuclear Laboratory, Durham, NC 27708, USA}
\newcommand{\utk}{Department of Physics and Astronomy, University of Tennessee, Knoxville, TN 37996, USA}
\newcommand{\ornl}{Oak Ridge National Laboratory, Oak Ridge, TN 37831, USA}
\newcommand{\sandia}{Sandia National Laboratories, Livermore, CA 94550, USA}
\newcommand{\fermi}{Enrico Fermi Institute and Kavli Institute for Cosmological Physics, University of Chicago, Chicago, IL 60637, USA}
\newcommand{\chicago}{Department of Physics, University of Chicago, Chicago, IL 60637, USA}
\newcommand{\nmsu}{Department of Physics, New Mexico State University, Las Cruces, NM 88003, USA}
\newcommand{\lanl}{Los Alamos National Laboratory, Los Alamos, NM, USA, 87545, USA}
\newcommand{\cenpa}{Center for Experimental Nuclear Physics and Astrophysics \& Department of Physics, University of Washington, Seattle, WA 98195, USA}
\newcommand{\ncsu}{Department of Physics, North Carolina State University, Raleigh, NC 27695, USA}
\newcommand{\usd}{Physics Department, University of South Dakota, Vermillion, SD 57069, USA}
\newcommand{\virgtech}{Center for Neutrino Physics, Virginia Tech, Blacksburg, VA 24061, USA}
\newcommand{\nccu}{Department of Mathematics and Physics, North Carolina Central University, Durham, NC 27707, USA}
\newcommand{\cmu}{Department of Physics, Carnegie Mellon University, Pittsburgh, PA 15213, USA}
\newcommand{\florida}{Department of Physics, University of Florida, Gainesville, FL 32611, USA}
\newcommand{\laurentian}{Department of Physics, Laurentian University, Sudbury, Ontario P3E 2C6, Canada}
\newcommand{\tufts}{Department of Physics and Astronomy, Tufts University, Medford, MA 02155, USA}
\newcommand{\kaist}{Department of Physics at Korea Advanced Institute of Science and Technology (KAIST), Daejeon, 34051, Republic of Korea}
\newcommand{\ibs}{Institute for Basic Science (IBS), Daejeon, 34051, Republic of Korea} 
\newcommand{\mipt}{Moscow Institute of Physics and Technology, Dolgoprudny, Moscow Region 141700, Russian Federation}

\author{D.~Akimov}\affiliation{\itep}\affiliation{\mephi}
\author{J.B.~Albert}\affiliation{\indiana}
\author{P.~An}\affiliation{\duke}\affiliation{\tunl}
\author{C.~Awe}\affiliation{\duke}\affiliation{\tunl}
\author{P.S.~Barbeau}\affiliation{\duke}\affiliation{\tunl}
\author{B.~Becker}\affiliation{\utk}
\author{V.~Belov}\affiliation{\itep}\affiliation{\mephi}
\author{I.~Bernardi}\affiliation{\utk}
\author{M.A.~Blackston}\affiliation{\ornl}
\author{L.~Blokland}\affiliation{\utk}
\author{A.~Bolozdynya}\affiliation{\mephi}
\author{B.~Cabrera-Palmer}\affiliation{\sandia}
%\author{M.~Cervantes}\affiliation{\duke}
\author{N.~Chen}\affiliation{\cenpa}
\author{D.~Chernyak}\affiliation{\usd}
\author{E.~Conley}\affiliation{\duke}
\author{R.L.~Cooper}\affiliation{\nmsu}\affiliation{\lanl}
\author{J.~Daughhetee}\affiliation{\utk}
\author{M.~del~Valle~Coello}\affiliation{\indiana}
\author{J.A.~Detwiler}\affiliation{\cenpa}
%\author{M.~D'Onofrio}\affiliation{\indiana}
\author{M.R.~Durand}\affiliation{\cenpa}
\author{Y.~Efremenko}\affiliation{\utk}\affiliation{\ornl}
%\author{E.M.~Erkela}\affiliation{\cenpa}
\author{S.R.~Elliott}\affiliation{\lanl}
\author{L.~Fabris}\affiliation{\ornl}
\author{M.~Febbraro}\affiliation{\ornl}
\author{W.~Fox}\affiliation{\indiana}
\author{A.~Galindo-Uribarri}\affiliation{\utk}\affiliation{\ornl}
\author{A. Gallo Rosso}\affiliation{\laurentian} 
\author{M.P.~Green}\affiliation{\tunl}\affiliation{\ornl}\affiliation{\ncsu}
\author{K.S.~Hansen}\affiliation{\cenpa}
\author{M.R.~Heath}\affiliation{\ornl}
\author{S.~Hedges}\affiliation{\duke}\affiliation{\tunl}
\author{M.~Hughes}\affiliation{\indiana}
\author{T.~Johnson}\affiliation{\duke}\affiliation{\tunl}
\author{M.~Kaemingk}\affiliation{\nmsu}
\author{L.J.~Kaufman}\altaffiliation[Now at: ] {SLAC National Accelerator Laboratory, Menlo Park, CA 94205, USA}\affiliation{\indiana}
\author{A.~Khromov}\affiliation{\mephi}
\author{A.~Konovalov}\affiliation{\itep}\affiliation{\mephi}
\author{E.~Kozlova}\affiliation{\itep}\affiliation{\mephi}
\author{A.~Kumpan}\affiliation{\mephi}
\author{L.~Li}\affiliation{\duke}\affiliation{\tunl}
\author{J.T.~Librande}\affiliation{\cenpa}
\author{J.M.~Link}\affiliation{\virgtech}
\author{J.~Liu}\affiliation{\usd}
\author{K.~Mann}\affiliation{\tunl}\affiliation{\ornl}
\author{D.M.~Markoff}\affiliation{\tunl}\affiliation{\nccu}
\author{O.~McGoldrick}\affiliation{\cenpa}
\author{H.~Moreno}\affiliation{\nmsu}
\author{P.E.~Mueller}\affiliation{\ornl}
\author{J.~Newby}\affiliation{\ornl}
\author{D.S.~Parno}\affiliation{\cmu}
\author{S.~Penttila}\affiliation{\ornl}
\author{D.~Pershey}\affiliation{\duke}
\author{D.~Radford}\affiliation{\ornl}
\author{R.~Rapp}\affiliation{\cmu}
\author{H.~Ray}\affiliation{\florida}
\author{J.~Raybern}\affiliation{\duke}
\author{O.~Razuvaeva}\affiliation{\itep}\affiliation{\mephi}
\author{D.~Reyna}\affiliation{\sandia}
\author{G.C.~Rich}\affiliation{\fermi}
\author{D.~Rudik}\affiliation{\itep}\affiliation{\mephi}
\author{J.~Runge}\affiliation{\duke}\affiliation{\tunl}
\author{D.J.~Salvat}\affiliation{\indiana}
\author{K.~Scholberg}\affiliation{\duke}
\author{A.~Shakirov}\affiliation{\mephi}
\author{G.~Simakov}\affiliation{\itep}\affiliation{\mephi}\affiliation{\mipt}
\author{G.~Sinev}\affiliation{\duke}
\author{W.M.~Snow}\affiliation{\indiana}
\author{V.~Sosnovtsev}\affiliation{\mephi}
\author{B.~Suh}\affiliation{\indiana}
\author{R.~Tayloe}\affiliation{\indiana}
\author{K.~Tellez-Giron-Flores}\affiliation{\virgtech}
\author{R.T.~Thornton}\affiliation{\indiana}\affiliation{\lanl}
\author{I.~Tolstukhin}\altaffiliation[Now at: ] {Argonne National Laboratory, Argonne, IL 60439, USA}\affiliation{\indiana}
\author{J.~Vanderwerp}\affiliation{\indiana}
\author{R.L.~Varner}\affiliation{\ornl}
\author{C.J.~Virtue}\affiliation{\laurentian}
\author{G.~Visser}\affiliation{\indiana}
\author{C.~Wiseman}\affiliation{\cenpa}
\author{T.~Wongjirad}\affiliation{\tufts}
\author{J.~Yang}\affiliation{\tufts}
\author{Y.-R.~Yen}\affiliation{\cmu}
\author{J.~Yoo}\affiliation{\kaist}\affiliation{\ibs}
\author{C.-H.~Yu}\affiliation{\ornl}
\author{J.~Zettlemoyer}\affiliation{\indiana}
\collaboration{COHERENT collaboration}

%% file: supplemental.tex
%\clearpage
\appendix*
\begin{figure*}
% \llap sequences put labels on figures
  \includegraphics[width=.45\textwidth]{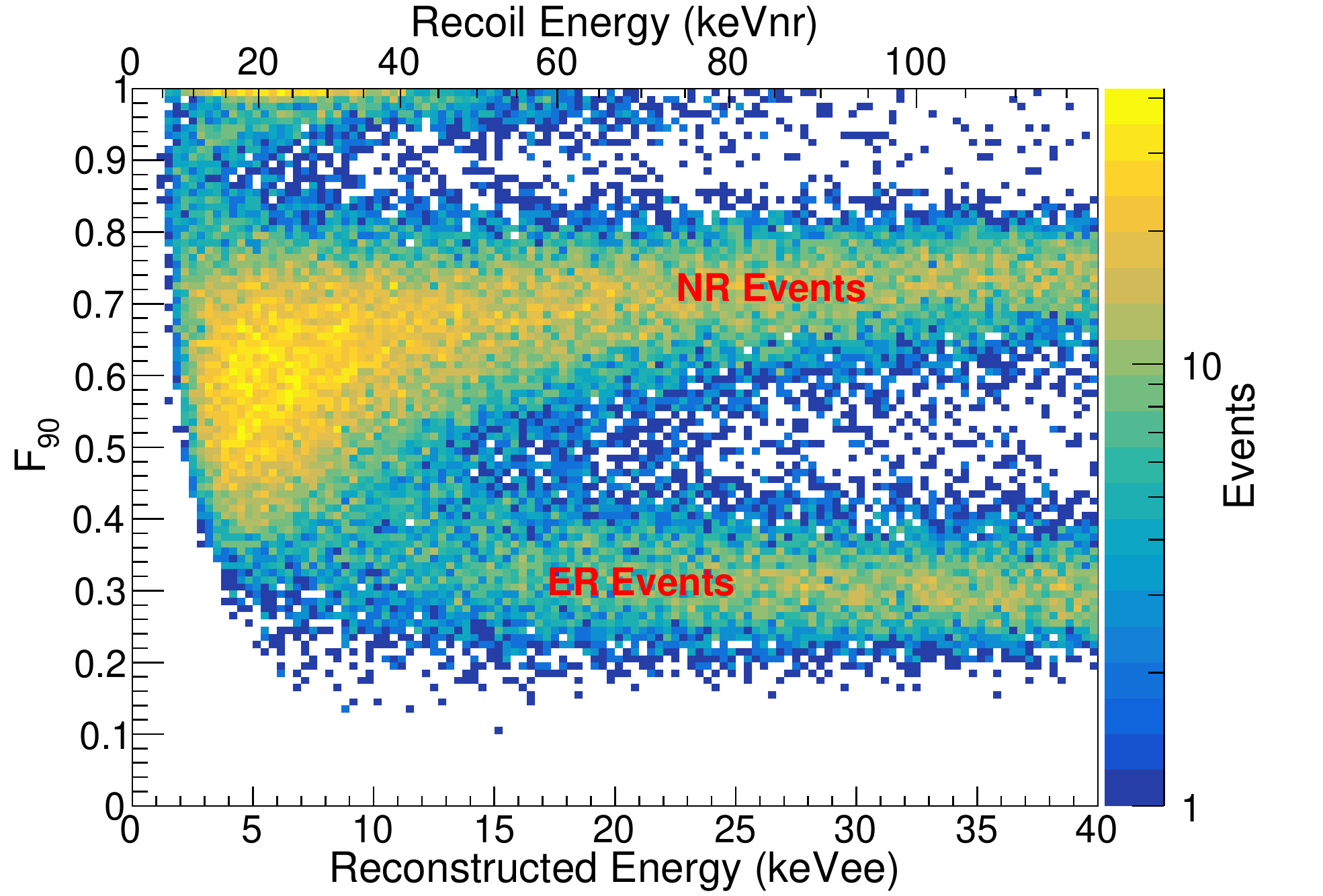}
    \llap{\parbox[b]{1.8in}{{\large a)}\\\rule{0ex}{0.35in}}}
  \includegraphics[width=.45\textwidth]{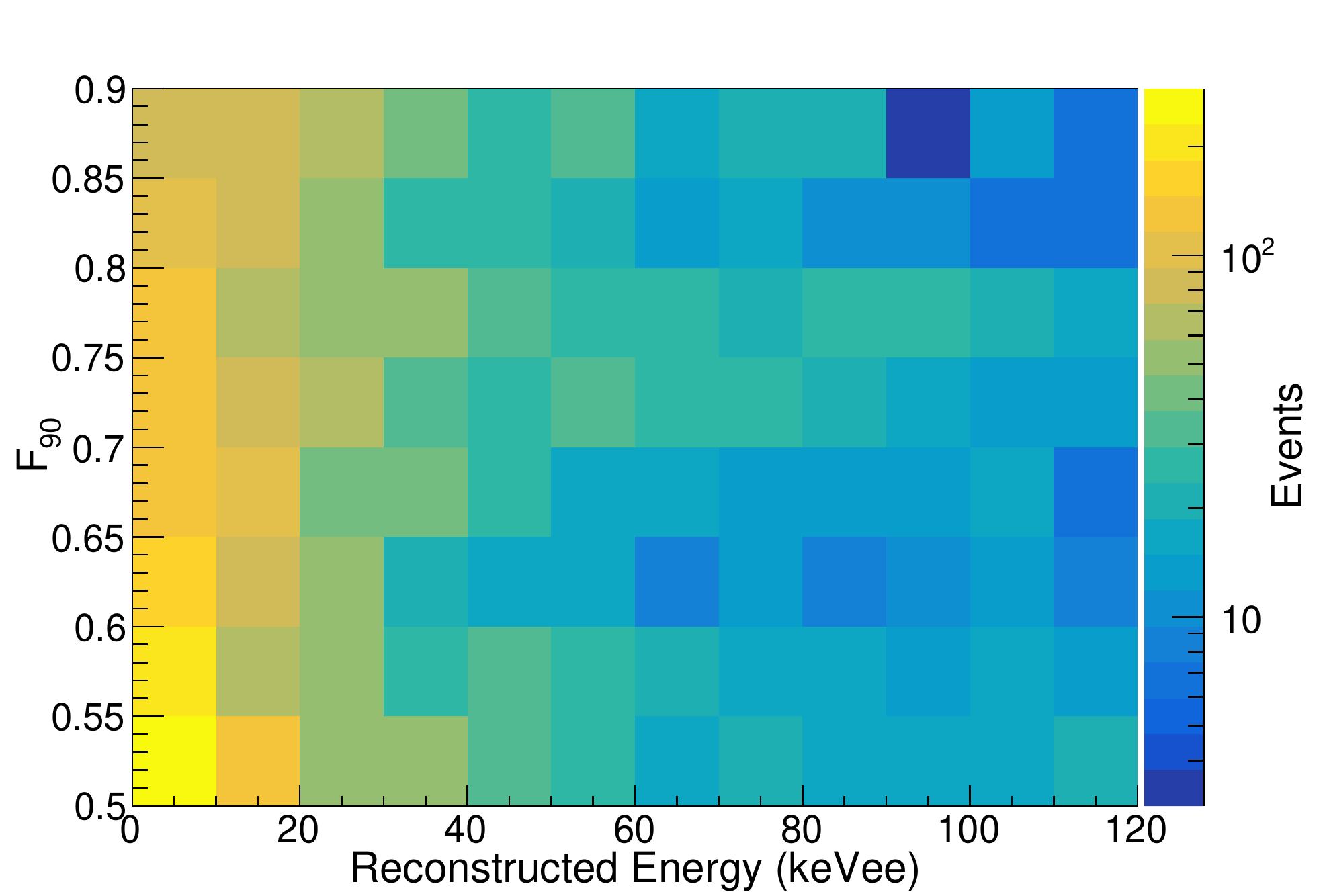} 
    \llap{\parbox[b]{1.8in}{{\large b)}\\\rule{0ex}{0.35in}}}
  \includegraphics[width=.45\textwidth]{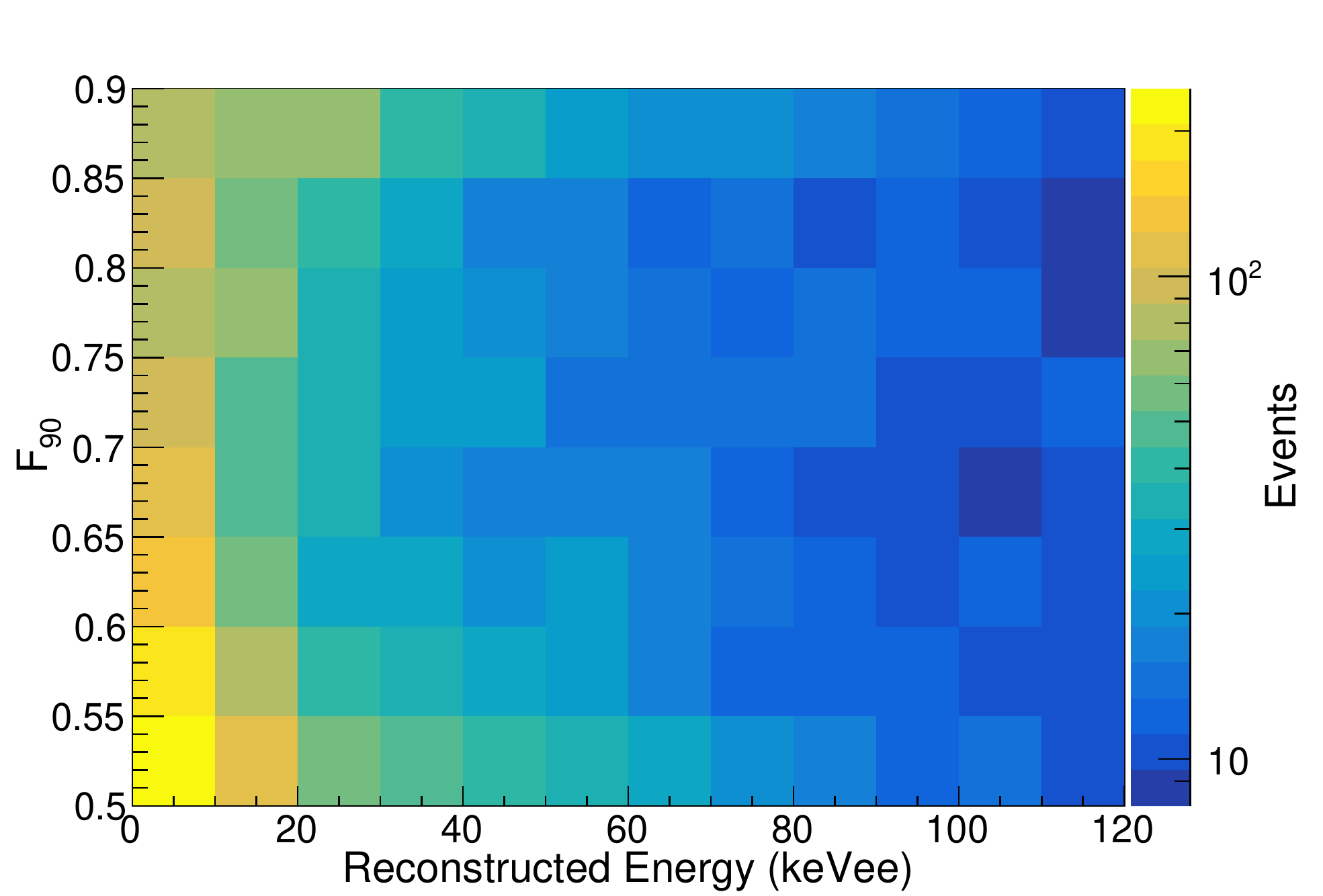}
    \llap{\parbox[b]{1.8in}{{\large c)}\\\rule{0ex}{0.35in}}}
  \includegraphics[width=.45\textwidth]{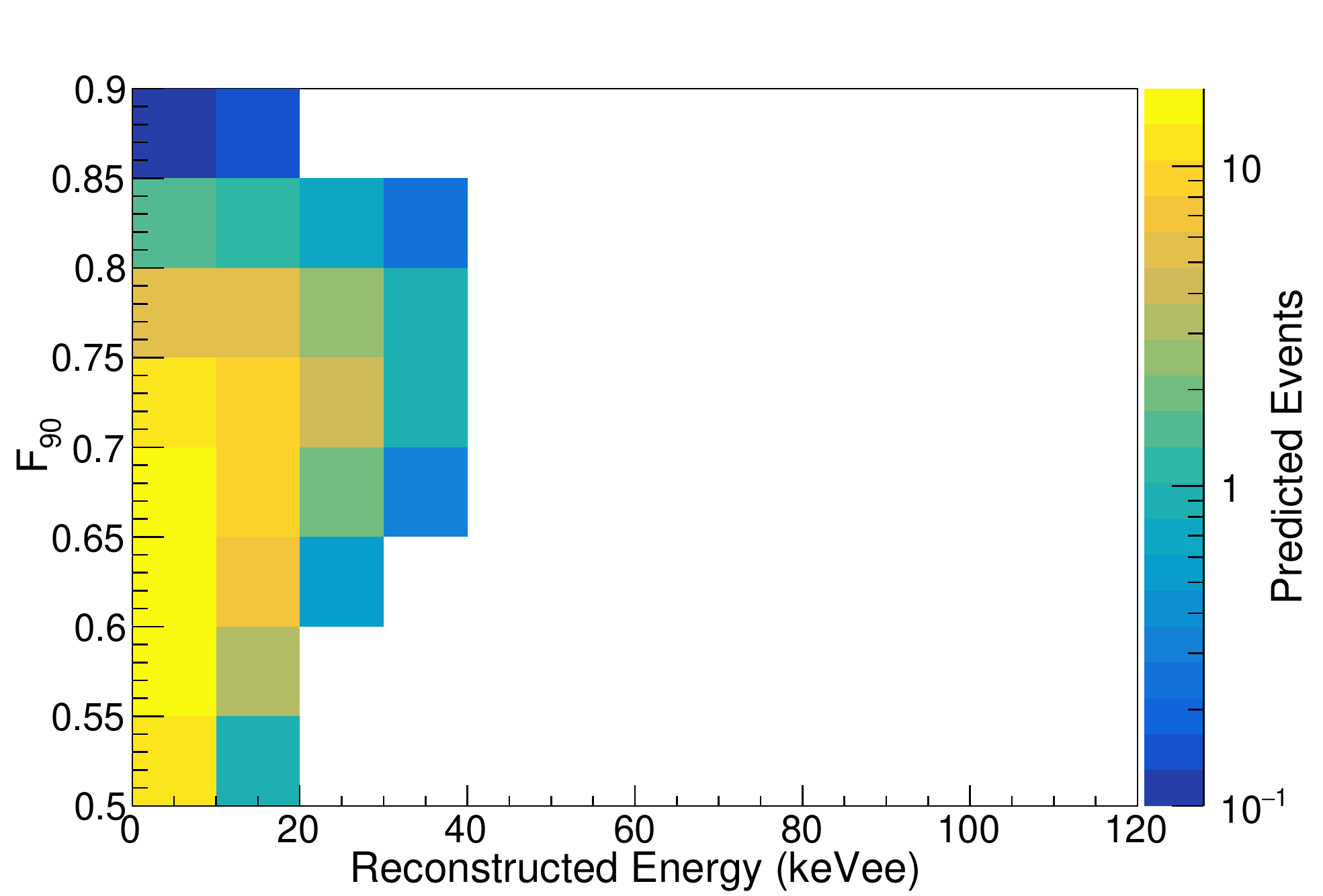}
    \llap{\parbox[b]{1.8in}{{\large d)}\\\rule{0ex}{0.35in}}}
  \caption{ Distribution of the PSD parameter \fninety vs. energy for an AmBe calibration source (a). NR events populate the band at $\fninety \approx 0.7$. The distributions are shown with different \fninety and energy ranges for data (b), predicted SS-background (c), and predicted \cevns signal (d). }
\label{fig:f90distributions}
\end{figure*}
\section{Appendix}
In this section, we present supplementary information supporting the material reported above. 
%\begin{center}
%\setcounter{page}{0}
%\pagenumbering{arabic}
%\setcounter{page}{1}
%Supplemental Material to: \\
%{\large\textbf{First Measurement of Coherent Elastic Neutrino-Nucleus Scattering on Argon}} \\
%(COHERENT collaboration)
%\end{center}

\subsection{Pulse shape discrimination}
The signal in this analysis is an argon nuclear recoil (NR) in liquid argon.  These NR events are separated from electron recoils (ER) via pulse-shape discrimination (PSD).  The PSD parameter, $\fninety=I_{90}/I$, is formed from the total number of photoelectrons, $I$, extracted from the corrected waveform in a 6~$\mu$s window after the initial pulse and the fast signal, $I_{90}$, dominated by singlet light, from the first $90$~ns of the pulse.

AmBe neutron source data are used to characterize the PSD response for NR events. The $\fninety$ distribution versus $E$ for these data as a function of energy is shown in Fig.~\ref{fig:f90distributions}a).  This AmBe source provides both neutrons and photons which explains the NR band at $\fninety \approx 0.7$ and the ER band at  $\fninety \approx 0.3$. The collection of events at $\fninety \approx 1.0$ are from beam-unrelated Cherenkov events. The behavior is broadly consistent with measurements from other LAr detectors~\cite{Hitachi:1983zz,Regenfus_2012}.  The mean value of $\fninety$ for both NR and ER events is parameterized versus energy for use in the simulation and construction of the expected distributions that are shown in Figs.~\ref{fig:f90distributions}b)-\ref{fig:f90distributions}d). 

\subsection{Liquid Argon Quenching Factor}
\begin{figure}
\includegraphics[width=0.9\columnwidth,page=9,trim={0 0 15 15},clip]{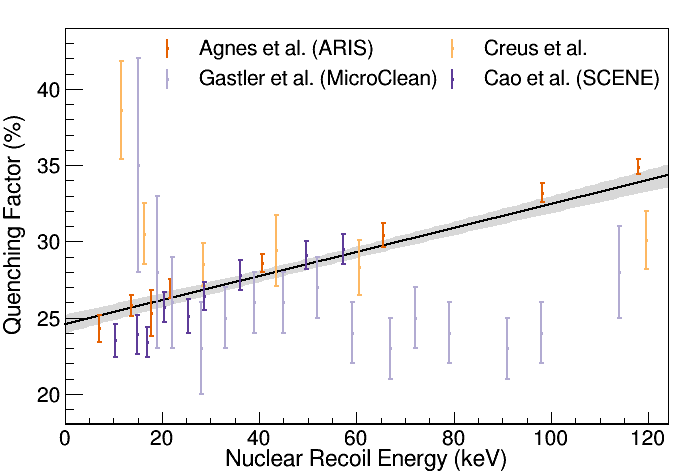}
\caption{\label{fig:larqf} Quenching factor fit with an error band constructed with a overall error scaling to yield $\chi^2/\mathrm{DOF}\approx 1$.}
\end{figure}

The predicted detector response to \cevns NR events relative to that for ER events is quantified with the so-called ``quenching factor'' \QF, defined as the ratio of the light output from NR events to that of ER events at the same kinetic energy. The most recent measurements of the \QF\ for liquid argon in the energy range $0$~--~$125$~keVnr~\cite{Agnes:2018mvl,Cao:2014gns,Creus:2015fqa,Gastler:2010sc} are shown in Figure~\ref{fig:larqf}. While there appears to be some tension between data sets, there is no \textit{a priori} reason to discard any particular measurement.  Therefore, we decided to do a simultaneous fit of these data with a complete treatment of the errors.  Since \cenns has little efficiency for $E<20$~keVnr, and any evidence for a more complex energy dependence is not clear, we assumed a linear model for the energy dependence of the \QF.

The simultaneous fit to this data utilized a standard least-squares method~\cite{PDG2018}, with an error matrix to handle any correlations in a particular data set.  While it is reasonable to assume that all individual data sets contained correlated error, only Ref.~\cite{Gastler:2010sc} reported them, so the error matrix contained off-diagonal terms only for these data. We suggest here in passing that future \QF\ measurements report correlated errors (as would occur, for example, with an overall energy calibration uncertainty) with the \QF\ data, allowing for a more correct treatment of errors in fits to world data.  The linear fit to this data with the reported errors yielded a $\chi^2/DOF$ of 138.1/36.  Following the recommended method of Ref.~\cite{PDG2018}, the errors on all data points were simultaneously scaled by a factor of 2.0 such that $\chi^2/DOF=1$.  This yields $\mathrm{QF}=(0.246 \pm 0.006) + ((7.8 \pm 0.9)\times10^{-4}\mathrm{~keVnr}^{-1})T$ with a correlation coefficient of -0.79 between the slope and intercept.  This fit and resulting error band are shown in Figure~\ref{fig:larqf}. A factor of 2 increase in the errors brings the data into reasonable agreement when correlated errors are considered.

A 2\% error on the efficiency for acceptance of \cevns events resulted from this \QF\ uncertainty and was calculated by varying the \QF\ used in the simulation within the error band of Figure~\ref{fig:larqf}.  Other scenarios for the energy dependence below 20~keVnr were also considered to quantify extreme possibilities.  If the data only from Refs.~\cite{Agnes:2018mvl,Cao:2014gns} (\cite{Creus:2015fqa,Gastler:2010sc}) are used for the fit of \QF\ below 20~keVnr, a change in the \cevns acceptance of -1\% (+12\%) results.  

\subsection{Maximum Likelihood Analysis}
\begin{figure}
\includegraphics[width=0.9\columnwidth, trim={0 0 40 30},clip]{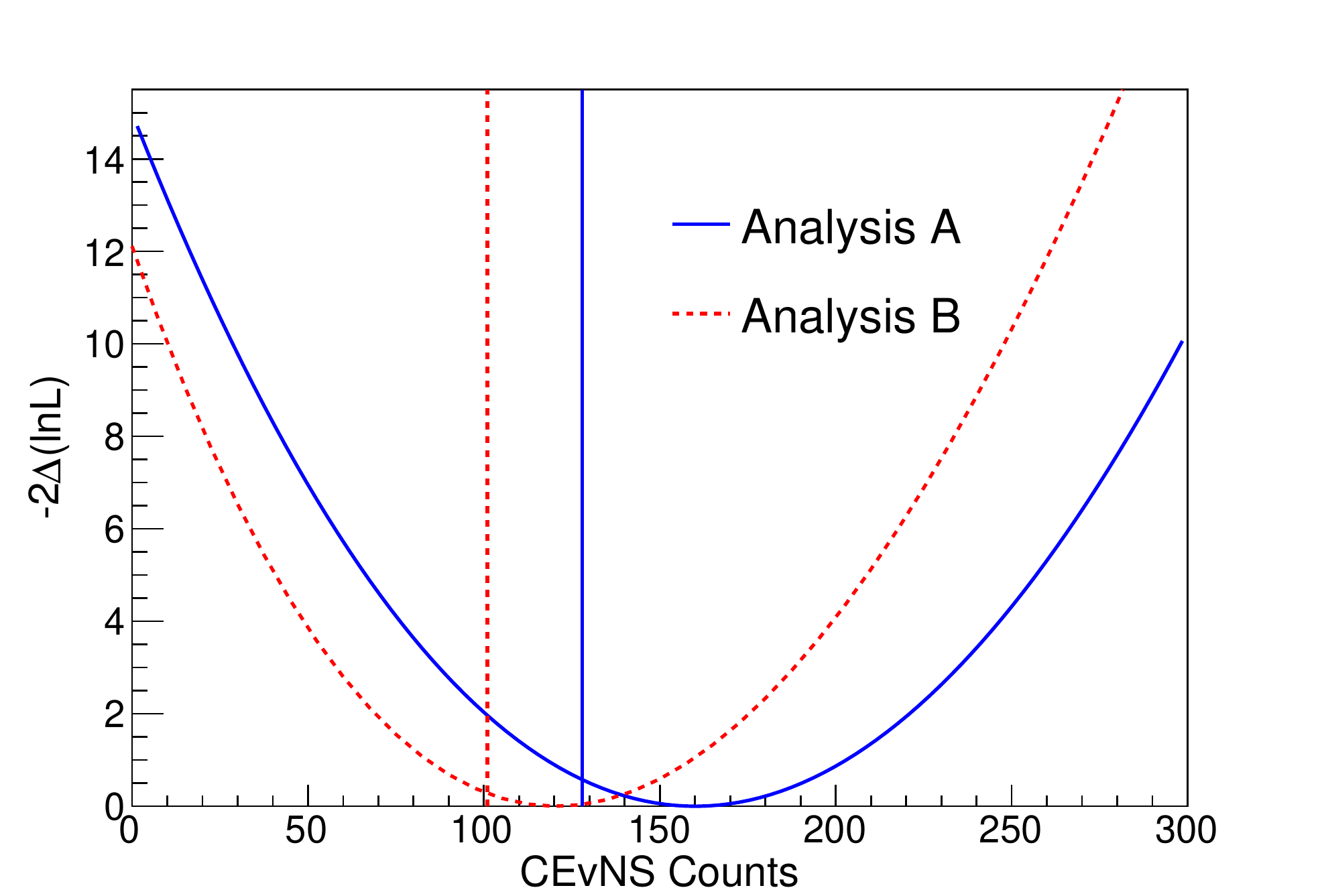}
\caption{\label{fig:profilell} The likelihood function (curves) vs predicted number of \cevns events profiled over the number of SS and BRN background events for both Analysis A and B. The vertical lines show the predicted number of events from the SM \cevns cross section accounting for detector response.
}
\end{figure}

\begin{figure*}
\begin{minipage}{0.325\textwidth}
\includegraphics[width=\textwidth,trim={0 1 40 0},clip]{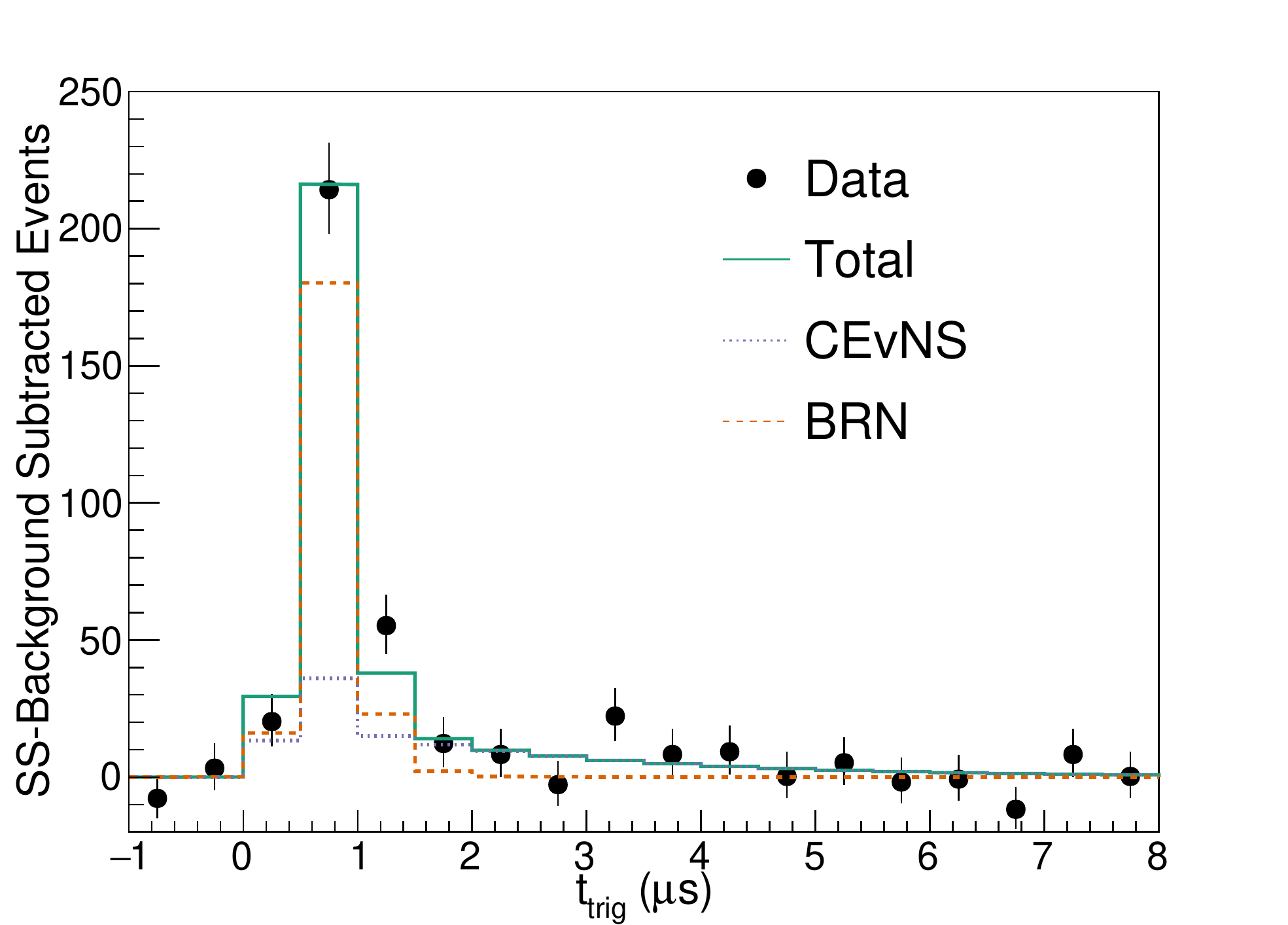}
\end{minipage}
\hfill
\begin{minipage}{0.325\textwidth}
\includegraphics[width=\textwidth,trim={0 1 40 0},clip]{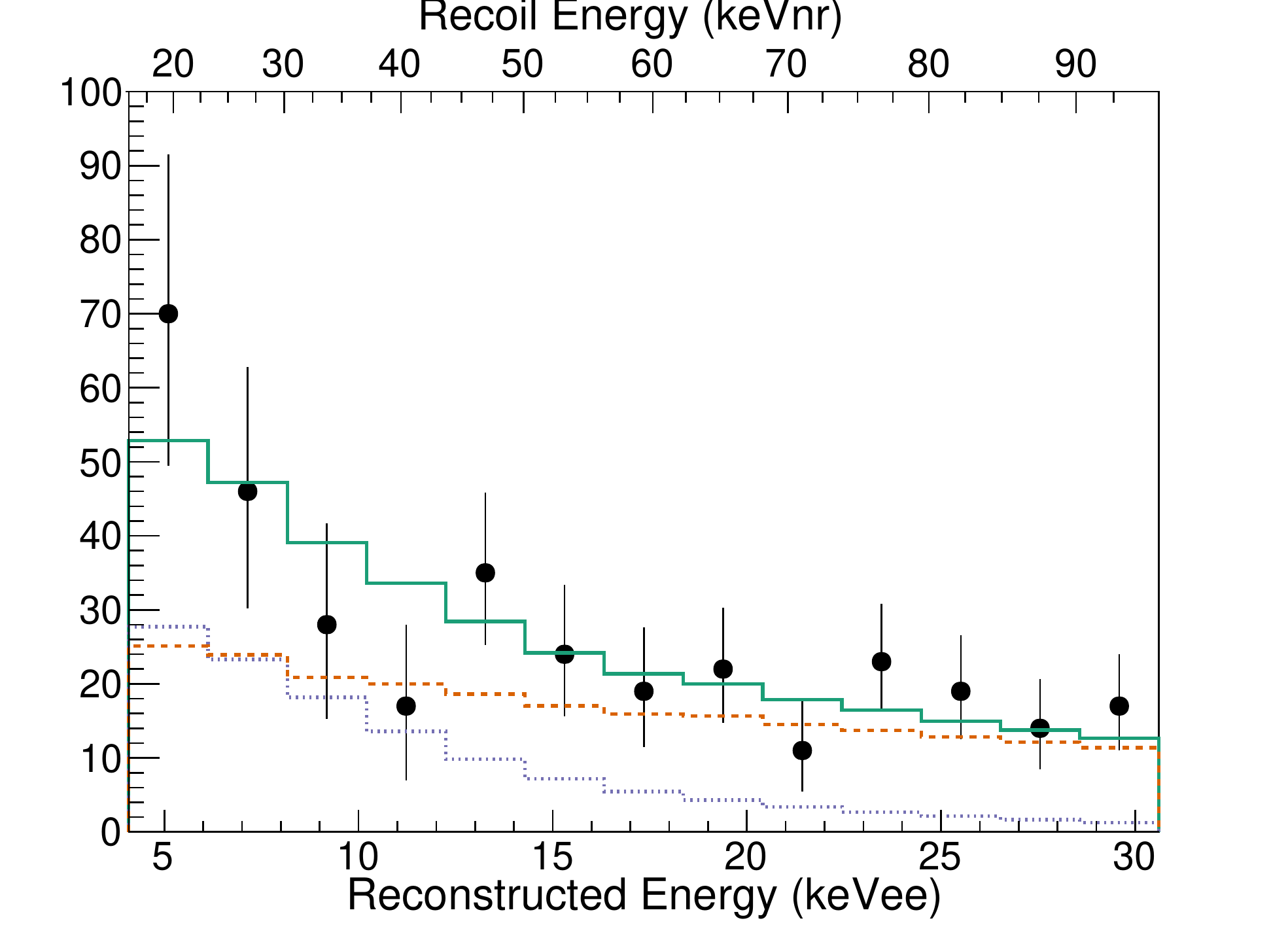}
\end{minipage}
\hfill
\begin{minipage}{0.325\textwidth}
\includegraphics[width=\textwidth,trim={0 1 40 0},clip]{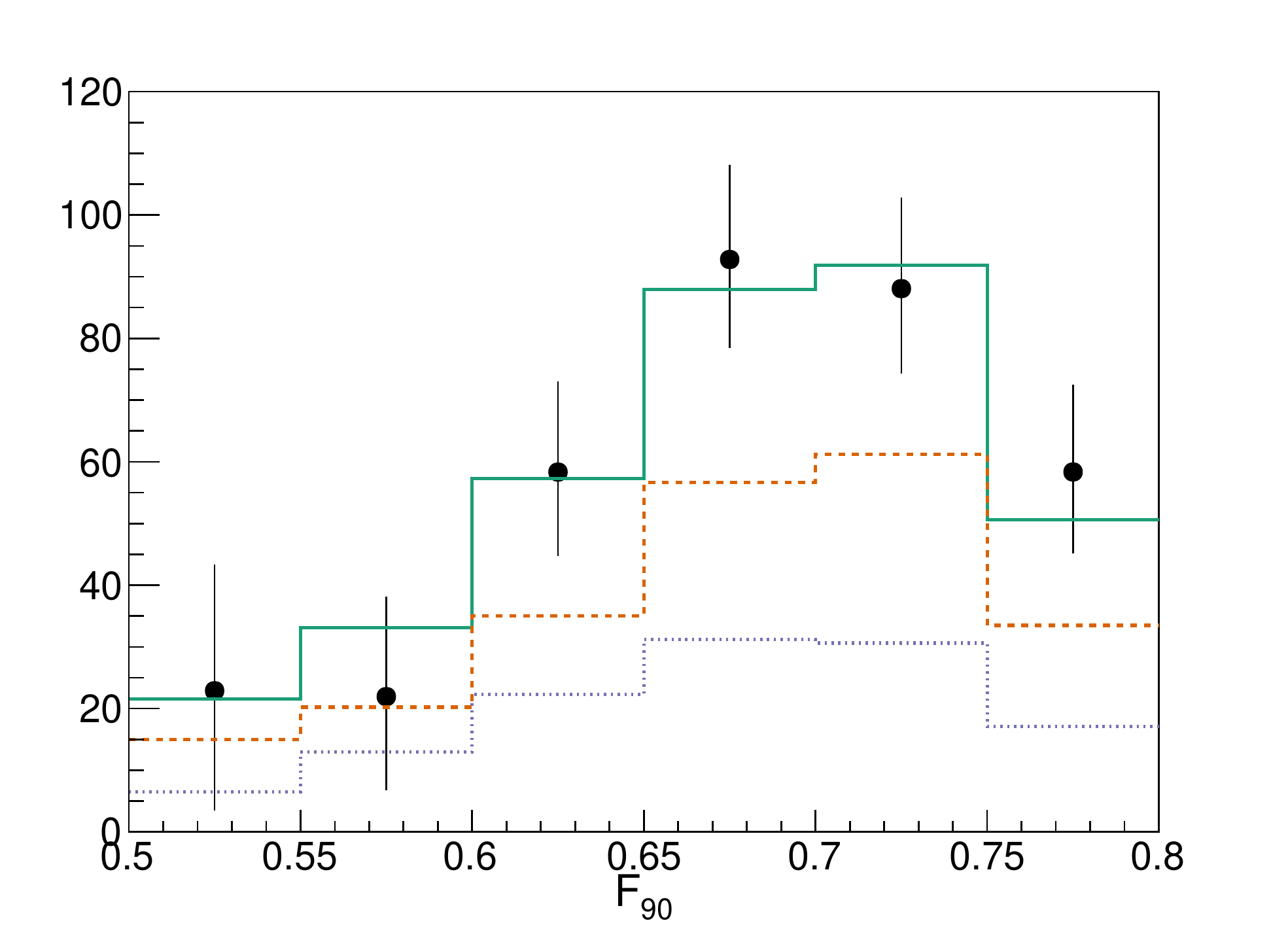}
\end{minipage}
\caption{\label{fig:AnalysisRprojs}
Projection of the maximum likelihood PDF from Analysis B on $t_\mathrm{trig}$ (left), reconstructed energy (center), and $\fninety$ (right).  The fit SS background has been subtracted to better show the \cevns component.   Bin-bin systematic errors were not calculated in this analysis.} 
\end{figure*}

\begin{figure*}
\begin{minipage}{0.325\textwidth}
\includegraphics[width=\textwidth,trim={0 1 40 0},clip]{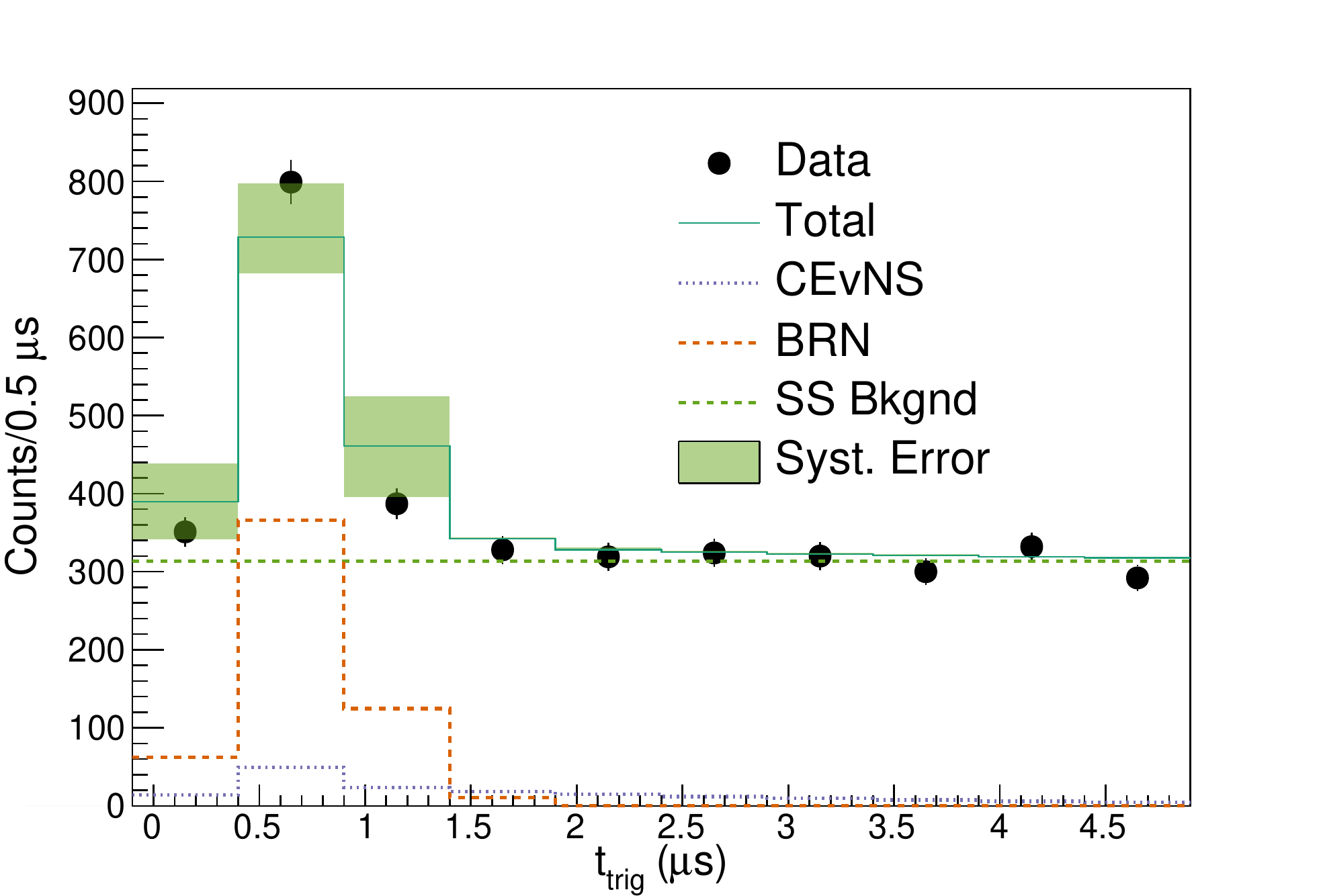}
\end{minipage}
\hfill
\begin{minipage}{0.325\textwidth}
\includegraphics[width=\textwidth,trim={0 1 40 0},clip]{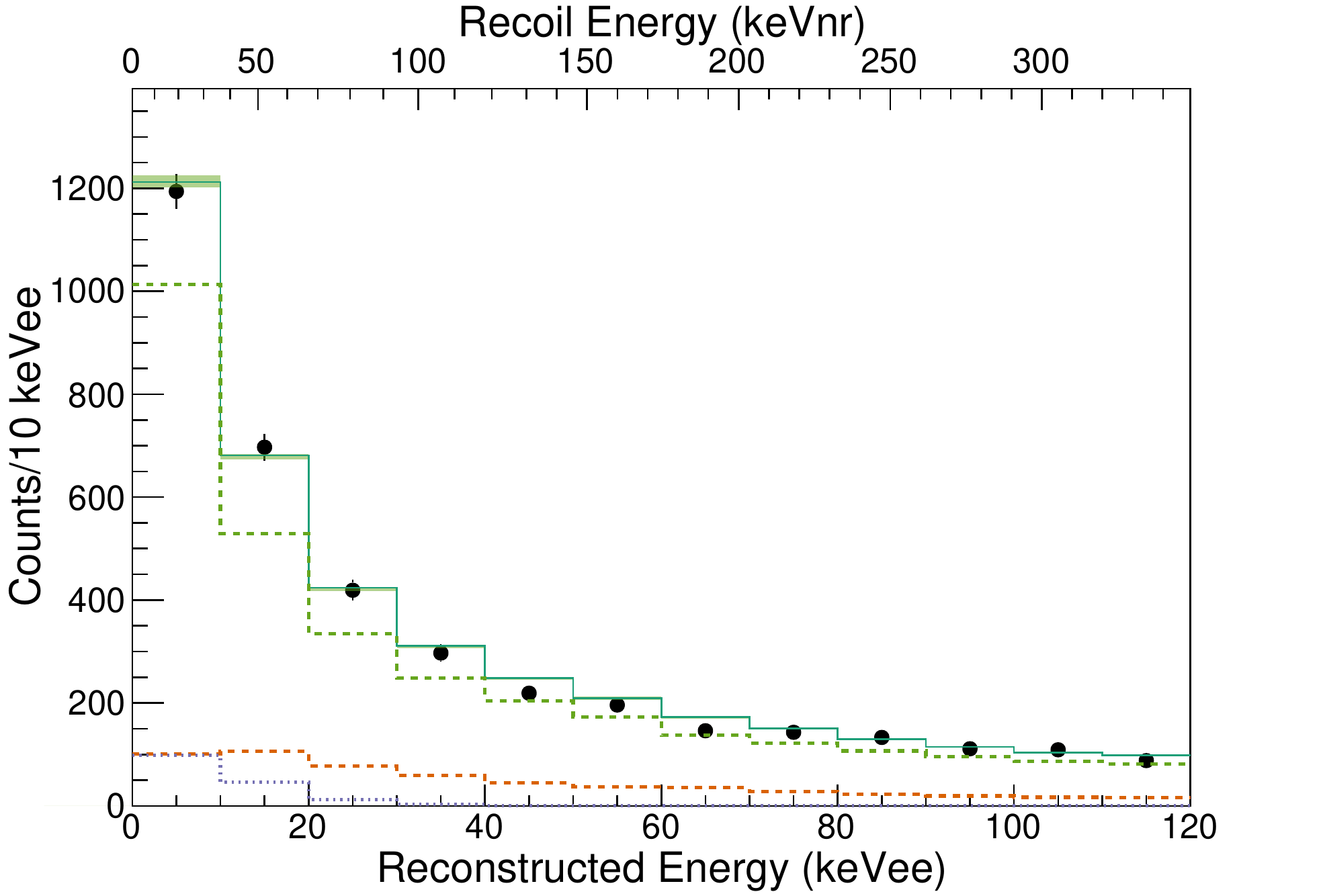}
\end{minipage}
\hfill
\begin{minipage}{0.325\textwidth}
\includegraphics[width=\textwidth,trim={0 1 40 0},clip]{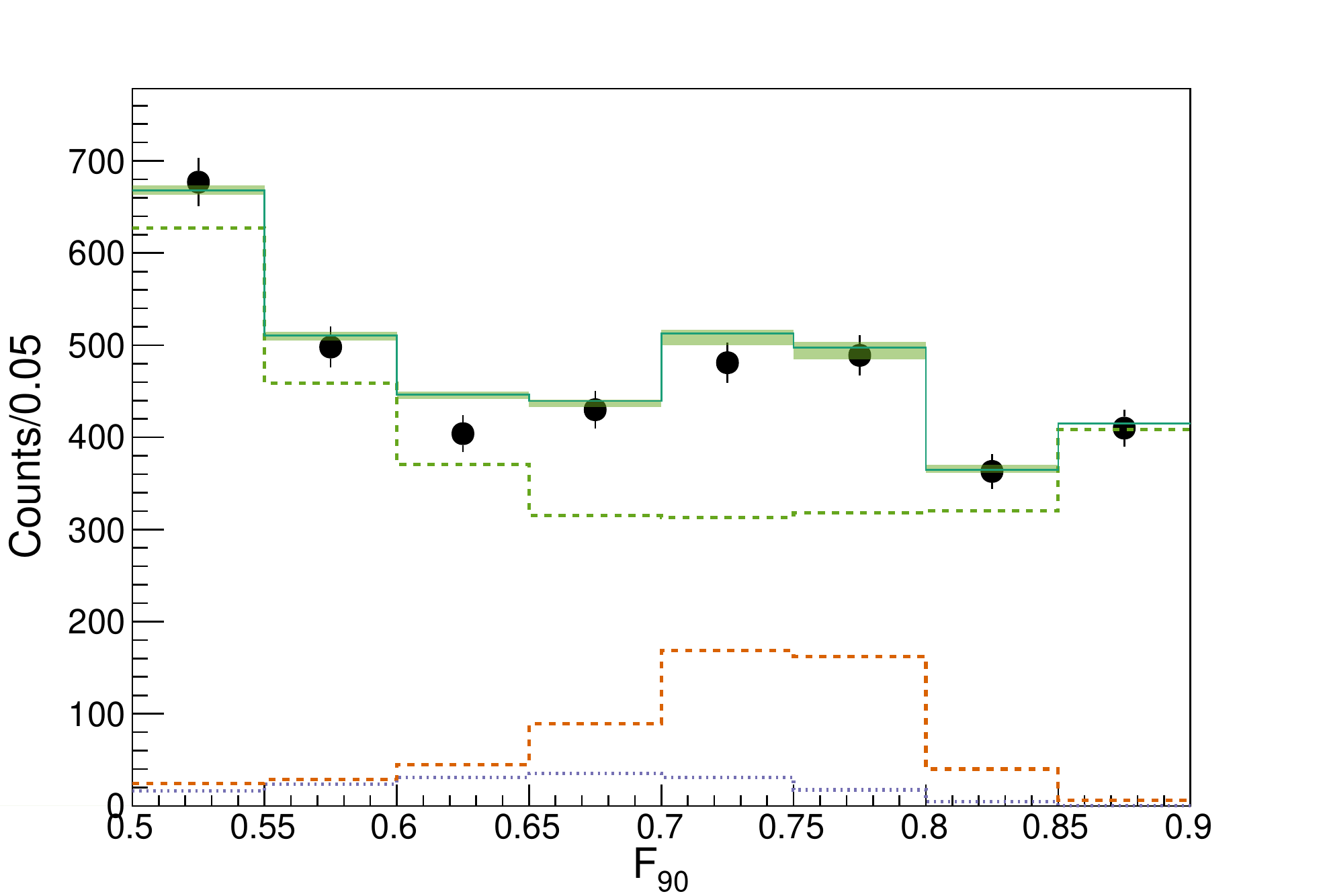}
\end{minipage}
\caption{\label{fig:unsubllprojections} Projection of the best-fit maximum likelihood probability density function (PDF) from Analysis A on $t_\mathrm{trig}$ (left), reconstructed energy (center), and $\fninety$ (right) along with selected data and statistical errors. The fit SS background is included in these projections. The green band shows the envelope of fit results resulting from the $\pm 1\sigma$ systematic errors on the PDF.}
\end{figure*}

A maximum likelihood fit was used to find the best estimate of $\Ncevns$ for the results reported here. The statistics-only null significance is determined by forming the quantity,
\begin{equation}
-2\Delta \ln L = -2(\ln L(\Ncevns)-\ln L_\mathrm{best}),
\end{equation}
that depends on the difference between the likelihoods at a given value of $\Ncevns$ and at the best-fit value of \Ncevns, $L_\mathrm{best}$.  The value of this quantity at $\Ncevns=0$ determines the statistics-only null-rejection significance with the assumption that it is distributed as a $\chi^2$ function with 1 degree of freedom. This assumption was tested with pseudo-data and supports our simple treatment of systematic errors in this analysis. Figure~\ref{fig:profilell} shows $-2\Delta \ln L$ profiled over the number of SS and BRN background events for the data sets in analyses A and B.
Figure~\ref{fig:AnalysisRprojs} shows the projections of the likelihood fit for analysis B.  Figure~\ref{fig:unsubllprojections} shows the projections of the likelihood fit for analysis A, but without subtraction of the SS background distribution.

\subsection{\cevns Cross Section $N$ Dependence}
\begin{figure}
\includegraphics[width=0.99\columnwidth,trim={0 0 40 30},clip]{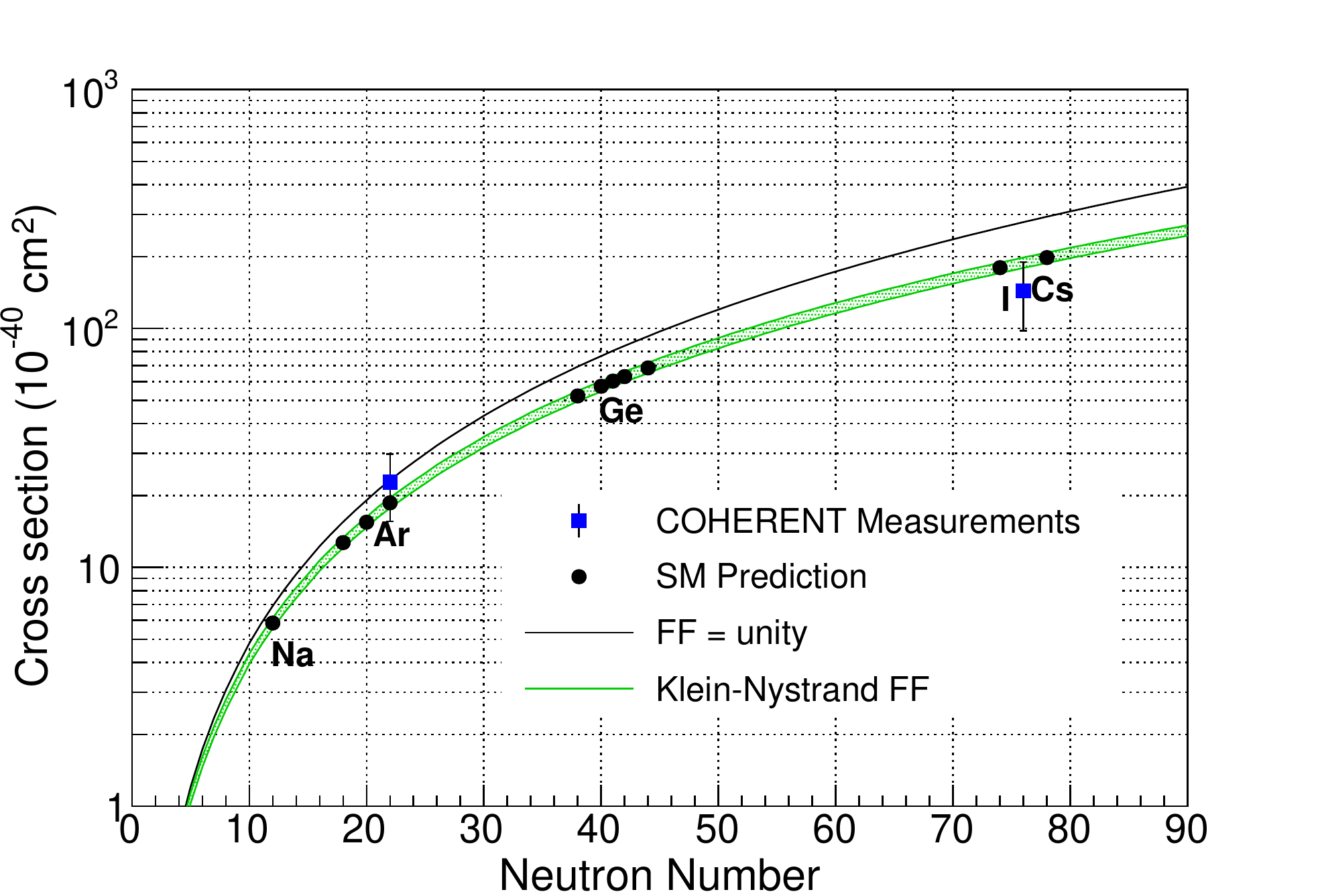} 
\caption{\label{fig:xsectionvN} The measured \cevns flux-weighted cross section from this analysis together with the previous results for CsI[Na]~\cite{Akimov:2017ade} and as expected in the SM as a function of neutron number. Expectations for planned COHERENT target nuclei are also computed. The form factor (FF) unity assumption is compared to the Klein-Nystrand~\cite{klein1999} value that is used for this analysis with the green band representing a $\pm3$\% variation on the neutron radius. 
}
\end{figure}

With the result reported here, the COHERENT collaboration has measured the flux-weighted \cevns cross section with different nuclei.  These results, along with the SM prediction, are shown in Figure~\ref{fig:xsectionvN}.
